\documentclass[letterpaper,11pt]{article}
\usepackage{epic,eepic,amsmath,latexsym,amssymb,color}
\usepackage{ifthen,graphics,epsfig,subfigure,fullpage}
\usepackage[english]{babel}
\usepackage{times}

\usepackage{array}
\usepackage{color}
\usepackage{colortbl}
\usepackage{multirow}
\usepackage{pgf}
\usepackage{tikz}
\usetikzlibrary{arrows,matrix, calc, fit}
\usepackage{ifthen}

\usepackage{rotating}

\usepackage{pgfplots}

\usepackage{comment}
\usepackage{caption}

\usepackage{intcalc}
\usetikzlibrary{patterns}
\usepackage{floatrow}
\usepackage{layout}

\newlength {\squarewidth}

\newcommand{\toto}{xxx}
\newenvironment{proofT}{\noindent{\bf
Proof }} {\hspace*{\fill}$\Box$\par\vspace{3mm}}
\newenvironment{proofL}{\noindent{\bf
Proof }} {\hspace*{\fill}$\Box$\par\vspace{3mm}}

\newcounter{linecounter}
\newcommand{\linenumbering}{\ifthenelse{\value{linecounter}<10}
{(0\arabic{linecounter})}{(\arabic{linecounter})}}
\renewcommand{\line}[1]{\refstepcounter{linecounter}\label{#1}\linenumbering}
\newcommand{\resetline}[1]{\setcounter{linecounter}{0}#1}
\renewcommand{\thelinecounter}{\ifnum \value{linecounter} >
9\else 0\fi \arabic{linecounter}}

\newcommand{\vir}[1]{``#1''}

\newcommand{\ceil}[1]{\lceil #1 \rceil}
\newcommand{\op}[1]{${\sf #1}()$}

\newcommand{\msg}[1]{{\sc #1}$()$}

\newcommand{\ang}[2]{\langle #1, #2 \rangle}

\newcommand{\Mod}[1]{\ (\text{mod}\ #1)}

\newcommand{\intervallo}[3]{ 
	\draw[-] (#1,-.5) -- (#1+\deltaPiccolo,-.5);	
	\draw [-] (#1+\deltaPiccolo,-.5) -- (#1+#2*\deltaPiccolo,-.5);	
	\foreach \x in {0,...,#2}
	\draw[] (#1+\x*\deltaPiccolo, \n) -- (#1+\x*\deltaPiccolo, -.6);
	\node[]() at (#1, -.75) {\scriptsize{$#3$}};
	\node[]() at (#1+#2*\deltaPiccolo+.1, -.75) {\scriptsize{$#3+#2\delta$}};
}

\newcommand{\punto}[4]{
	\fill[#3] (#1,#2) circle[radius=2pt];
	\node[] () at (#1,#2-.2) {#4};
}

\newcommand{\lineaVerticale}[4]{
	\draw[#4] (#1, #2) -- (#1, \n+.5);
	\node[] () at (#1, #2-.2) {#3};
}

\newcommand{\rettangoloOpaco}[4]{
	\filldraw[fill=#4!40!white, fill opacity=0.1, draw=black] (#1,#2) rectangle (#1+#3,#2-.4);
}

\newcommand{\curatoParziale}[4]{
	\filldraw[fill=#4!40!white, draw=black] (#1,#2) rectangle (#1+#3,#2-.2);
}

\newcommand{\processes}{ 
	\foreach \x in {0,...,\n} 
	\draw[->] (0,\x) -- (\lenght,\x); 
	\foreach \x in {0,..., \n}
	\node[]() at (-.3,\x) {$s_ \x$};	
}

\newcommand{\faults}[5]{ 
	\foreach \x in {0,...,#3}
	\filldraw[fill=red!40!white, draw=black] (#1+\x*\deltaGrande,\intcalcMod{ #2+\x}{\n+1}) rectangle (#1+\deltaGrande+\x*\deltaGrande,\intcalcMod{ #2+\x}{\n+1}-.3);
	
	\foreach \x in {0,...,\intcalcSub{#3}{1}}
	\filldraw[fill=#4!40!white, draw=black] (#1+\deltaGrande+\x*\deltaGrande,\intcalcMod{#2+\x}{\n+1}) rectangle (#1+\deltaGrande+\x*\deltaGrande+\gammaCuring,\intcalcMod{#2+\x}{\n+1}-.2);
	
	\ifnum#5>0 
	\foreach \x in {0,...,#5}
		\filldraw[fill=#4!40!white, draw=black] (0,\intcalcMod{#2-\x+\n}{\n+2}) rectangle (\gammaCuring-\x*\deltaGrande,\intcalcMod{#2-\x+\n}{\n+2}-.2);	
	\fi
}

\newcommand{\faultsAdaptive}[5]{ 
	\foreach \x in {0,...,#3}
	
		\ifthenelse{{#1+\deltaGrande+\x*\deltaGrande = #5}}
			{\filldraw[fill=#4!40!white, draw=black] (#1+\x*\deltaGrande,\intcalcMod{#2+\x}{\n+1}) rectangle (#1+\deltaGrande+\x*\deltaGrande,\intcalcMod{#2+\x}{\n+1}+.3);}
			{\ifthenelse{{#1+\deltaGrande+\x*\deltaGrande < #5}}
				{\filldraw[fill=#4!40!white, draw=black] (#1+\x*\deltaGrande,\intcalcMod{#2+\x}{\n+1}) rectangle (#1+\deltaGrande+\x*\deltaGrande,\intcalcMod{#2+\x}{\n+1}+.3);}
				{\filldraw[fill=#4!40!white, draw=black] (#1+\x*\deltaGrande,\intcalcMod{#2+\x}{\n+1}) rectangle (#5,\intcalcMod{#2+\x}{\n+1}+.3);}}
	\foreach \x in {0,...,#3}
	\filldraw[fill=gray!40!white, draw=black] (#1+\deltaGrande+\x*\deltaGrande,\intcalcMod{#2+\x}{\n+1}) rectangle (#1+\deltaGrande+\x*\deltaGrande+\gammaCuring,\intcalcMod{#2+\x}{\n+1}+.2);
}

\newtheorem{definition}{Definition}
\newtheorem{lemma}{Lemma}
\newtheorem{corollary}{Corollary}
\newtheorem{theorem}{Theorem}

\title{Optimal Storage under \\Unsynchronized Mobile Byzantine Faults}

\author{\IEEEauthorblockN{Silvia Bonomi\IEEEauthorrefmark{1}, 
        Antonella Del Pozzo\IEEEauthorrefmark{1,2},
        Maria Potop-Butucaru\IEEEauthorrefmark{2}, 
        and S\'{e}bastien Tixeuil\IEEEauthorrefmark{2}}
      \IEEEauthorblockA{\IEEEauthorrefmark{1}Sapienza Universit\`{a} di Roma,Via Ariosto 25, 00185 Roma, Italy
        \IEEEauthorrefmark{2}UPMC Sorbonne Universit\'{e}s}
      }

\begin{document}

\title{\bf   Optimal Storage under Unsynchrononized Mobile Byzantine Faults}
\author{Silvia Bonomi$^\star$, Antonella Del Pozzo$^\star$$^\dagger$, Maria Potop-Butucaru$^\dagger$, S\'ebastien Tixeuil$^\dagger$\\~\\
$^\star$Sapienza Universit\`{a} di Roma,Via Ariosto 25, 00185 Roma, Italy\\
\texttt{\{bonomi, delpozzo\}}$@$dis.uniroma1.it\\
$^\dagger$Universit\'e Pierre \& Marie Curie (UPMC) -- Paris 6, France\\
\texttt{\{maria.potop-butucaru, sebastien.tixeuil\}}$@$lip6.fr}

\date{}
\maketitle
\thispagestyle{empty}

\begin{abstract}
In this paper we prove lower and matching upper bounds for the number of servers required to implement a regular shared register that tolerates \emph{unsynchronized Mobile Byzantine failures}. We consider the strongest model of Mobile Byzantine failures to date: agents are moved arbitrarily by an omniscient adversary from a server to another in order to deviate their computation in an unforeseen manner. When a server is infected by an Byzantine agent, it behaves arbitrarily until the adversary decides to \vir{move} the agent to another server. Previous approaches considered asynchronous servers with synchronous mobile Byzantine agents (yielding impossibility results), and synchronous servers with synchronous mobile Byzantine agents (yielding optimal solutions for regular register implementation, even in the case where servers and agents periods are decoupled). 

We consider the remaining open case of synchronous servers with unsynchronized agents, that can move at their own pace, and change their pace during the execution of the protocol. Most of our findings relate to lower bounds, and characterizing the model parameters that make the problem solvable. It turns out that unsynchronized mobile Byzantine agent movements requires completely new proof arguments, that can be of independent interest when studying other problems in this model. Additionally, we propose a generic server-based algorithm that emulates a regular register in this model, that is tight with respect to the number of mobile Byzantine agents that can be tolerated. Our emulation spans two awareness models: servers with and without self-diagnose mechanisms. In the first case servers are aware that the mobile Byzantine agent has left and hence they can stop running the protocol until they recover a correct state while in the second case, servers are not aware of their faulty state and continue to run the protocol using an incorrect local state.

\end{abstract}

\section{Introduction}

Byzantine fault tolerance is a fundamental building block in distributed system, as Byzantine failures include all possible faults, attacks, virus infections  and  arbitrary behaviors that can occur in practice (even unforeseen ones). The classical setting considers Byzantine participants remain so during the entire execution, yet software rejuvenation techniques increase the possibility that a  corrupted node \emph{does not remain corrupted during the whole system execution} and may be aware of its previously compromised status \cite{SBAM09}.     


Mobile Byzantine Failures (MBF) models have been recently introduced to integrate those concerns. Then, faults are represented by Byzantine agents that are managed by an omniscient adversary that ``moves'' them from a host process to another, an agent being able to corrupt its host in an unforeseen manner. MBF investigated so far consider mostly \emph{round-based} computations, and can be classified according to Byzantine mobility constraints: \emph{(i)} constrained mobility~\cite{Garay+95+ORA} agents may only move from one host to another when protocol messages are sent (similarly to how viruses would propagate), while \emph{(ii)} unconstrained mobility~\cite{Banu+2012,BDNP14,Garay+1994,Ostrovsky+91,Reischuk+85,Sasaki+2013} agents may move independently of protocol messages. In the case of unconstrained mobility, several variants were investigated~\cite{Banu+2012,BDNP14,Garay+1994,Ostrovsky+91,Reischuk+85,Sasaki+2013}: Reischuk~\cite{Reischuk+85} considers that malicious agents are stationary for a given period of time, Ostrovsky and Yung~\cite{Ostrovsky+91} introduce the notion of mobile viruses and define the adversary as an entity that can inject and distribute faults; finally, Garay~\cite{Garay+1994}, and more recently Banu \emph{et al.}~\cite{Banu+2012}, and Sasaki \emph{et al.}~\cite{Sasaki+2013} and Bonnet  \emph{et al.} \cite{BDNP14} consider that processes execute synchronous rounds composed of three phases: \emph{send}, \emph{receive}, and \emph{compute}. Between two consecutive such synchronous rounds, Byzantine agents can move from one node to another. Hence the set of faulty hosts at any given time has a bounded size, yet its membership may evolve from one round to the next. The main difference between the aforementioned four works~\cite{Banu+2012,BDNP14,Garay+1994,Sasaki+2013} lies in the knowledge that hosts have about their previous infection by a Byzantine agent. In Garay's model~\cite{Garay+1994}, a host is able to detect its own infection after the Byzantine agent left it. Sasaki \emph{et al.}~\cite{Sasaki+2013} investigate a model where hosts cannot detect when Byzantine agents leave. Finally, Bonnet \emph{et al.}~\cite{BDNP14} considers an intermediate setting where cured hosts remain in \emph{control} on the messages they send (in particular, they send the same message to all destinations, and they do not send obviously fake information, \emph{e.g.} fake id). Those subtle differences on the power of Byzantine agents turns out to have an important impact on the bounds for solving distributed problems.

A first step toward decoupling algorithm rounds from mobile Byzantine moves is due to Bonomi \emph{et al.}~\cite{BDPT16c}. In their model, mobile Byzantine movements are either: \emph{(i)} synchronized, but the period of movement is independent to that of algorithm rounds, \emph{(ii)} independent time bounded, meaning that Byzantine agents are only requested to remain some minimum amount of time at any occupied node, or \emph{(iii)} independent time unbounded, which can be seen as a special case of \emph{(ii)} when the minimum amount of time is one time unit. In particular, the Bonomi \emph{et al.}~\cite{BDPT16c} model implies that Byzantine moves are no more related to messages that are exchanged through the protocol. 

\paragraph{Register Emulation.} Traditional solutions to  build a Byzantine tolerant storage service (\emph{a.k.a.} register emulation) can be divided into two categories: \emph{replicated state machines} \cite{S90}, and \emph{Byzantine quorum systems} \cite{B00,MR98,MAD02,MAD02-2}. Both approaches are based on the idea that the current state of the storage is replicated among processes, and the main difference lies in the number of replicas that are simultaneously involved in the state maintenance protocol. Several works investigated the emulation of  self-stabilizing or pseudo-stabilizing Byzantine tolerant  SWMR or MWMR registers \cite{AADDPT15,BPT15,BDPR15}. All these works do not consider the complex case of mobile Byzantine faults.
Recently, Bonomi \emph{et al.}~\cite{BDP16} proposed optimal self-stabilizing atomic register implementations for round-based synchronous systems under the four Mobile Byzantine models described in \cite{Banu+2012,BDNP14,Garay+1994,Sasaki+2013}. The round-free model~\cite{BDPT16c} where Byzantine moves are decoupled from protocol rounds also enables optimal solutions (with respect to the number of Byzantine agents) for the implementation of regular registers. However, this last solution requires Byzantine agents to move in synchronous steps, whose duration for the entire execution is fixed, so the movements of Byzantine agents is essentially synchronous. As it is impossible to solve the register emulation problem when processes are asynchronous and Byzantine agents are synchronous~\cite{BDPT16c}, the only case remaining open is that of synchronous processes and unsynchronized Byzantine agents.  

\paragraph{Our Contribution.} We relax the main assumption made for obtaining positive results in the round-free model: Byzantine moves are no more synchronized.
The main contribution of this paper is to thoroughly study
the impact of \emph{unsynchronized} mobile Byzantine agents on the register emulation problem. We present lower and matching upper bounds for implementing a regular register in the unsynchronized mobile Byzantine model. We first explore and characterize the key parameters of the model that enable problem solvability. As expected, the lower bounds results require completely new proof techniques that are of independent interest while studying other classical problems in the context of unsynchronized mobile Byzantine agents. When the problem is solvable, it turns out that minor changes to existing quorum-based protocols joint with smart choices of quorums thresholds command optimal resilience (with respect to the number of Byzantine agents). Table \ref{tab:allBounds} summarizes all the lower bounds for the various models, the newly obtained results are presented in boldface.


\begin{table*}[]
\caption{Summary of lower bounds in different system models. $\delta$ is the upper bound on the message delay, and $\Delta$ is the period for synchronized agent moves (in the synchronous agents setting) or the lower bound for an agent to remain on a server (in the unsynchronized agents setting).}
\label{tab:allBounds}
\hspace{-6cm}
\begin{minipage}[b]{0.32\hsize}
\begin{tabular}{|c|c|c|c|}
\hline
\multicolumn{4}{|l|}{Round-based model \cite{BDP16}} \\ \hline
Burhman   & Garay   & Bonnet  & Sasaki  \\ \hline
$2f+1$    & $3f+1$  & $4f+1$  & $4f+1$  \\ \hline
\end{tabular}
\end{minipage}\hspace{1.5cm}
\begin{minipage}[b]{0.32\hsize}
\begin{tabular}{l|l|l|l|l|}
\cline{2-5}
                                                      & \multicolumn{4}{l|}{Round-free model}                                \\ \hline
\multicolumn{1}{|l|}{Agents moves}                & \multicolumn{2}{l|}{Synchronized} & \multicolumn{2}{l|}{\bf Unsynchronized} \\ 
\multicolumn{1}{|l|}{}                & \multicolumn{2}{l|}{\cite{BDPT16c}} & \multicolumn{2}{l|}{\bf [this paper]} \\ 
\hline
\multicolumn{1}{|l|}{Cured state awareness}               & Aware          & Unaware         & {\bf Aware}           & {\bf Unaware}         \\ \hline
\multicolumn{1}{|l|}{$\delta \leq \Delta < 2\delta$}  & $5f+1$            & $8f+1$           & \boldmath$6f+1$            & \boldmath$12f+1$            \\ \hline
\multicolumn{1}{|l|}{$2\delta \leq \Delta < 3\delta$} & $4f+1$            & $5f+1$           & \boldmath$4f+1$            & \boldmath$7f+1$            \\ \hline
\end{tabular}
\end{minipage}
\end{table*}


\section{System Model }\label{sec:systemModel}

We consider a distributed system composed of an arbitrary large set of client processes $\mathcal{C}$, and a set of $n$ server processes $\mathcal{S}=\{s_1, s_2 \dots s_n\}$. Each process in the distributed system (\emph{i.e.}, both servers and clients) is identified by a unique identifier. Servers run a distributed protocol emulating a shared memory abstraction, and clients are unaware of the protocol run by the servers.
The passage of time is measured by a fictional global clock (\emph{e.g.}, that spans the set of natural integers), whose processes are unaware of.
At each time instant $t$, each process (either client or server) is characterized by its \emph{internal state}, \emph{i.e.}, by the set of its local variables and their assigned values.
We assume that an arbitrary number of clients may crash, and that up to $f$ servers host, at any time $t$, a Byzantine agent. 
Furthermore, servers processes execute the same algorithm, and \emph{cannot} rely on high level primitives such as consensus or total order broadcast.  

\noindent{\bf Communication model.} 
Processes communicate through message passing. In particular, we assume that: \emph{(i)} each client $c_i \in \mathcal{C}$ can communicate with every server through a ${\sf broadcast}()$ primitive, \emph{(ii)} each server can communicate with every other server through a ${\sf broadcast}()$ primitive, and \emph{(iii)} each server can communicate with a particular client through a ${\sf send}()$ unicast primitive. We assume that communications are authenticated (\emph{i.e.}, given a message $m$, the identity of its sender cannot be forged) and reliable (\emph{i.e.}, spurious messages are not created and sent messages are neither lost nor duplicated).

\noindent{\bf Timing Assumptions.} The system is \emph{round-free synchronous} in the sense that: \emph{(i)} the processing time of local computations (except for {\sf wait} statements) are negligible with respect to communication delays, and are assumed to be equal to $0$, and \emph{(ii)}  messages take time to travel to their destination processes. In particular, concerning point-to-point communications, we assume that if a process sends a message $m$ at time $t$ then it is delivered by time $t+ \delta_p$ (with $\delta_p >0$). Similarly, let $t$ be the time at which a process $p$ invokes the ${\sf broadcast}(m)$ primitive, then there is a constant $\delta_b$ (with $\delta_b \ge \delta_p$) such that all servers have delivered $m$ at time $t+\delta_b$. For the sake of presentation, in the following we consider a unique message delivery delay $\delta$ (equal to $\delta_b \ge \delta_p$), and assume $\delta$ is known to every process.

\noindent{\bf Computation model.} Each process of the distributed system executes a distributed protocol $\mathcal{P}$ that is composed by a set of distributed algorithms. Each algorithm in $\mathcal{P}$ is represented by a finite state automaton and it is composed of a sequence of computation and communication steps. A computation step is represented by the computation executed locally to each process while a communication step is represented by the sending and the delivering events of a message. Computation steps and communication steps are generally called \emph{events}.
\begin{definition}[Execution History]
Let $\mathcal{P}$ be a distributed protocol.
Let $H$ be the set of all the events generated by $\mathcal{P}$ at any process $p_i$ in the distributed system and let $\rightarrow$ be the happened-before relation.
An execution history (or simply history) $\hat{H}= (H, \rightarrow)$ is a partial order on $H$ satisfying the relation $\rightarrow$.
\end{definition}

\begin{definition}[Valid State at time $t$]
Let $\hat{H}= (H, \rightarrow)$ be an execution history of a generic computation and let $\mathcal{P}$ be the corresponding protocol.
Let $p_i$ be a process and let $state_{p_i}$ be the state of $p_i$ at some time $t$.
$state_{p_i}$ is said to be valid at time $t$ if it can be generated by executing $\mathcal{P}$ on $\hat{H}$.
\end{definition}



\noindent\textbf{MBF model.}
We now recall the generalized Mobile Byzantine Failure model~\cite{BDPT16c}. Informally, in the MBF model, when a Byzantine agent is hosted by a process, the agent takes entire control of its host making it Byzantine faulty (\emph{i.e.}, it can corrupt the host's local variables, forces it to send arbitrary messages, etc.). Then, the Byzantine agent leaves its host with a possible corrupted state (that host is called \emph{cured}) before reaching another host.
We assume that any process previously hosting a Byzantine agent has access to a tamper-proof memory storing the correct protocol code. However, a cured server may still have a corrupted internal state, and thus cannot be considered correct.
The moves of a Byzantine agent are controlled by an omniscient adversary.

\begin{definition}[Correct process at time $t$]
Let $\hat{H}= (H, \rightarrow)$ be a history, and let $\mathcal{P}$ be the protocol generating $\hat{H}$.
A process is \emph{correct at time $t$} if \emph{(i)} it is correctly executing $\mathcal{P}$, and \emph{(ii)} its state is valid at time $t$.
We denote by $Co(t)$ the set of correct processes at time $t$. Given a time interval $[t, t']$, we denote by $Co ([t, t'])$ the set of all processes that remain correct during $[t, t']$ (\emph{i.e.}, $Co ([t, t'])= \bigcap_{\tau ~\in ~[t, t']} Co(\tau)$).
\end{definition}

\begin{definition}[Byzantine process at time $t$]
Let $\hat{H}= (H, \rightarrow)$ be a history, and let $\mathcal{P}$ be the protocol generating $\hat{H}$.
A process is \emph{Byzantine at time $t$} if it is controlled by a Byzantine agent and does not execute $\mathcal{P}$. 
We denote by $B(t)$ the set of Byzantine processes at time $t$. Given a time interval $[t, t']$, we denote by $B ([t, t'])$ the set of all processes that remain Byzantine during $[t, t']$ (\emph{i.e.}, $B ([t, t'])= \bigcap_{\tau ~\in ~[t, t']} B(\tau)$).
\end{definition}

\begin{definition}[Cured process at time $t$]
Let $\hat{H}= (H, \rightarrow)$ be a history, and let $\mathcal{P}$ be the protocol generating $\hat{H}$.
A process is \emph{cured at time $t$} if \emph{(i)} it is correctly executing $\mathcal{P}$, and \emph{(ii)} its state is not valid at time $t$.
We denote by $Cu(t)$ the set of cured processes at time $t$. Given a time interval $[t, t']$, we denote by $Cu ([t, t'])$ the set of all processes that remain cured during $[t, t']$ (\emph{i.e.}, $Cu ([t, t'])= \bigcap_{\tau ~\in ~[t, t']} Cu(\tau)$).
\end{definition}

With respect to the {\em movements} of agents, we consider the 
{\bf independent time-bounded (ITB)} model: each mobile Byzantine agent $ma_i$ is forced to remain on a host for at least a period $\Delta_i$. Given two mobile Byzantine Agents $ma_i$ and $ma_j$, their movement periods $\Delta_i$ and $\Delta_j$ may be different.
Note that previous results considering decoupled Byzantine moves~\cite{BDPT16c} were established in the weaker $\Delta$-synchronized model, where the external adversary moves \emph{all} controlled mobile Byzantine agents at the same time $t$, and their movements happen periodically with period $\Delta$. None of those properties remain valid in our model. 

Concerning the \emph{knowledge} that each process has about its failure state, we distinguish the following two cases:
{\bf Cured Aware Model (CAM)}: at any time $t$, every process is aware about its failure state; 
{\bf Cured Unaware Model (CUM)}: at any time $t$, every process is not aware about its failure state.


We assume that the adversary can control at most $f$ Byzantine agents at any time (\emph{i.e.}, Byzantine agents do not replicate while moving). In our work, only servers can be affected by the mobile Byzantine agents\footnote{It is trivial to prove that in our model, if clients are Byzantine, it is impossible to implement deterministically even a safe register. A Byzantine client may always introduce a corrupted value, and a server cannot distinguish between a correct client and a Byzantine one.}. It follows that, at any time $t$ $|B(t)|\le f$. However, during the system lifetime, all servers may be hosting a Byzantine agent at some point (\emph{i.e.}, none of the servers is guaranteed to remain correct forever).

\noindent{\bf Register Specification.}
\label{sec:Problem}

A register is a shared variable accessed by a set of processes, called clients, through two operations, namely ${\sf read}$ and ${\sf write}$. Informally, the ${\sf  write}$ operation updates the value stored in the shared variable, while  the $\sf read$ obtains the  value contained in the variable (\emph{i.e.}, the last written value). Every operation issued on a register is, generally, not instantaneous and it  can  be characterized  by  two events  occurring  at  its boundaries:  an \emph{invocation} event and a \emph{reply} event. These events occur at two time instants (called the invocation  time and the reply time) according  to the fictional global time.
  
An operation $op$ is \emph{complete} if both the  invocation event and the reply event occurred, otherwise, it \emph{failed}.
Given two operations  $op$ and $op'$,  their  invocation times ($t_{B}(op)$ and $t_B(op')$) and reply times ($t_E(op)$ and $t_E(op')$), $op$ \emph{precedes} $op'$ ($op \prec op'$) if and only if $t_E(op) < t_B(op')$. If $op$ does not precede $op'$ and $op'$  does not precede $op$, then $op$  and $op'$ are \emph{concurrent} (noted $op||op'$). Given a ${\sf write}(v)$ operation, the value $v$ is said to be written when the operation is complete. 

In this paper, we consider a single-writer/multi-reader (SWMR) regular register, as defined by Lamport~\cite{L86}, which is specified as follows:\\
 --- ${\sf Termination}$: if a correct client invokes an operation $op$, $op$ completes. \\
 --- ${\sf Validity}$: A $\sf{read}$ returns the last written value before its invocation (\emph{i.e.} the value written by the latest completed $\sf{write}$ preceding it), or a value written by a $\sf{write}$ concurrent with it.
 

\section{Lower bounds}

In this section we prove lower bounds with respect to the minimum fraction of correct servers to implement safe  registers in presence of mobile Byzantine failures \footnote{Results on safe register can be directly extended to the other register specifications.}. In particular we first prove lower bounds for the $(\Delta S, CAM)$ and $(\Delta S, CUM)$ models and then we extend those results to $(ITB, CAM)$ and $(ITB, CUM)$ models.
The first observation that raises is that in presence of mobile agents in the round-free models there are several parameters to take into account with respect to the round-based model. Let us start considering that the set of Byzantine servers changes its composition dynamically time to time. This yields to the following question: does it impact on the \op{read} duration? Or, in other words, such operation has to last as less as possible or until it eventually terminates? In this chapter we consider the \op{read} operation duration as a parameter itself, allowing us to easily verify when the variation of such parameter has any impact on lower bounds. Here below the list of parameters we take into account.
\begin{itemize}
\item servers knowledge about their failures state ($CAM, CUM$);
\item the relationship between $\delta$ and $\Delta$ (that states how many Byzantine servers there may be during an operation);
\item $T_r$, the \op{read} operations duration;
\item $\gamma$, the upper bound on the time during which a server can be in a cured state (the design of an optimal \op{maintenance} operation is out of the scope of this thesis, thus we use such upper bound as another parameter). 
\end{itemize}
Those parameters allow us to describe different failure models and help us to provide a general framework that produces lower bounds for each specific instance of the MBF models. In the sequel it will be clear that $\gamma$ varies depending on the coordinated/uncoordinated mobile agents movements ($\Delta S, ITB, ITU$). In other words, in this parameter is hidden the movements model taken into account, so we do not need to explicitly parametrize it.
Before to start let us precise that we do not consider the following algorithm families: (i) full information algorithm families (processes exchange information at each time instant); (ii) algorithms characterized by a read operation that does not require a request-reply pattern; (iii) algorithms with non quiescent operation (the message exchange triggered by an operation eventually terminates); and finally (iv) algorithms where clients interact with each other. All results presented in the sequel consider a families of algorithms such that previous characteristics do not hold.   
The lower bounds proof leverages on the classical construction of two indistinguishable executions. The tricky part is to characterize the set of messages delivered by a client from correct and incorrect servers depending of the ${\sf read}()$ operation duration. Let $T_r$, $T_r\geq 2\delta$ be such duration, each \op{read} operation requires at least a request-reply pattern). We first characterize the correct and incorrect sets of messages, delivered during $T_r$ time, with respect to $\Delta$ and $\gamma$.
For clarity, in the sequel we note \emph{correct  message/request/reply} a message that carries a valid value when it is sent (\emph{i.e.,} sent by a correct process). Otherwise, the message is \emph{incorrect}. 
It has been proven~\cite{BDPT16c} that a protocol $\mathcal{P}_{reg}$
implementing a regular register in a mobile Byzantine setting must
include in addition to the mandatory ${\sf read}$ and ${\sf write}$ operations an additional operation, ${\sf maintenance}$, defined below.

\begin{definition}[Maintenance operation ${\sf maintenance}$]
A ${\sf maintenance}$ operation is an operation that, when executed by a process $p_i$, terminates at some time $t$ leaving $p_i$ with a valid state at time $t$  (\emph{i.e.}, it guarantees that $p_i$ is correct at time $t$).
\end{definition}

Such operation has a direct impact on the number of correct processes in any time instant. For that reason it is important to characterize its duration, in particular its upper bound in terms of time. The following definition defines $\gamma$, the upper bound of the time during which a server can be in a cured state.

\begin{definition}[Curing time, $\gamma$]
We define $\gamma$ as the maximum time a server can be in a cured state. 
More formally, let $T_c$ the time at which server $s_c$ is left by a mobile agent, let $op_M$ the first ${\sf maintenance}$ operation that correctly terminates, then  $t_E(op_M)-T_c \leq \gamma$. 
\end{definition}

In order to build our indistinguishable execution, we define below a scenario of agents movement. Then, with respect this scenario, we construct two indistinguishable executions.

\begin{definition}[Scenario $S^*$]
 Let  $S^*$ be the following scenario: for each time $T_i,i\geq 0$ the affected servers are $s_{(i \mod{n})f+1}, \dots, s_{(i \mod{n})f+f}$. 
\end{definition}

\begin{figure*}
		\begin{tikzpicture}[y=-1cm]
		
		\def \lenght {5.3}
		\def \n {3} 
		
		\def \deltaGrande {0.6}
		\def \deltaPiccolo{0.5}
		\def \gammaCuring{1}
		

		\foreach \x in {0,...,\n} 
		\draw[->] (0,\x) -- (\lenght,\x); 
		\foreach \x in {0,..., \n}
		\node[]() at (-.3,\x) {\scriptsize{${\intcalcMod{\x+1}{\n+2}}f$}};

		\foreach \x in {0,...,7}
		\node[]() at (0+\x*\deltaGrande, -.6) {\scriptsize{$T_\x$}};
		
		\node[]() at (0+8*\deltaGrande, -.6) {\scriptsize{$\dots$}};
		
		\foreach \x in {0,...,7}
		\draw[dotted] (0+\x*\deltaGrande, \n) -- (0+\x*\deltaGrande, -.3);

		\faults{0}{0}{7}{gray}{1}
		\node[] () at (5.1,2.85) {$\dots$};			
		\end{tikzpicture}
		\caption{Representation of  $S^*$ where mobile agents affect groups of $f$ different servers each $T_i$ period. In particular here $\gamma > \Delta$.}\label{fig:scenario}
\end{figure*}
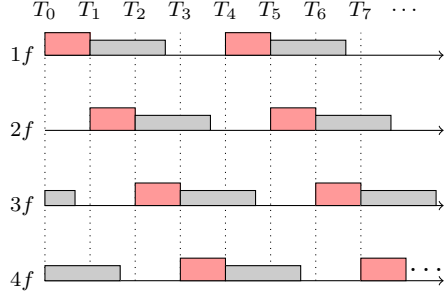

In Figure \ref{fig:scenario} is depicted $S^*$. In particular, the red part is the time where $f$ agents are affecting $f$ servers and the gray part is the time servers are running the ${\sf maintenance}$ operation. 

Let us characterize the $\mathcal{P}_{reg}$ protocol in the most general possible way. By definition a register abstraction involves ${\sf read}()$ and ${\sf write}()$ operations issued by clients. A ${\sf read}$ operation involves at least a ${\sf request-reply}$ communication pattern (i.e., two communication steps). Thus, given the system synchrony, a ${\sf read}()$ operation $op_R$ lasts at least $T_r \geq 2\delta$ time. Moreover we consider that a correct server sends a ${\sc reply}$ message in two occasions: \emph{(i)} after the delivery of a ${\sc request}$ message, and \emph{(ii)} right after it changes its state, at the end of the ${\sf maintenance}$ operation if an $op_R$ is occurring. The latter case exploits the ${\sf maintenance}$ operation allowing servers to reply with a valid value in case they were Byzantine at the beginning of the ${\sf read}$ operation. Moreover we assume that in $(*, CAM)$ model servers in a cured state do not participate to the ${\sf read}$ operation. Notice that those servers are aware of their current cured state and are aware of their impossibility to send correct replies. Even though those may seems not very general assumptions, let us just consider that we are allowing servers to correctly contribute to the computation as soon as they can and stay silent when they can not and under those assumptions we prove lower bounds. Thus if we remove those assumptions the lower bounds do not decreases. Scenario and protocol has been characterized. Now we aim to characterize the set of servers, regarding their failure states, that can appear during the execution of the protocol, in particular during the \op{read} operation. Those sets allow us to characterize correct and incorrect messages that a client delivers during a \op{read} operation.

\begin{definition}[Failure State of servers in a time interval]
Let $[t,t+T_t]$ be a time interval and let $t'$, $t'>0$, be a time instant. Let $s_i$ be a server and $state_i$ be $s_i$ state, $state_i \in \{correct,$ $ cured, Byzantine\}$. Let $S(t')$ be the set of servers $s_i$ that are in the state $state_i$ at $t'$, $S(t') \in \{Co(t'), Cu(t'), B(t')\}$.  $\tilde{S}(t,t+T_r)$ is the set of servers that have been in the state $state_i$ for at least one time unit during $[t,t+T_r]$.
More formally,  $\tilde{S}(t,t+T_r)= \bigcup_{t \le t' \le t+T_r} S(t')$.
\end{definition}

\begin{definition}[$\tilde{CBC}(t,t+T_r)$]
	Let $[t,t+T_r]$ be a time interval, $\tilde{CBC}(t,t+T_r)$ denotes servers that during a time interval $[t,t+T_r]$ belong first to $\tilde{B}(t,t+T_r)$ or $Cu(t)$ (only in $(\Delta S, CUM)$ model) and then to $Co(t+\delta,t+T_r-\delta)$ or vice versa.\\
	In particular let us denote:
	\begin{itemize}
	\item $\tilde{BC}(t,t+T_r)$ servers that during a time interval $[t,t+T_r]$ belong to $\tilde{B}(t,t+T_r)$ or $Cu(t)$ (only in $(\Delta S, CUM)$ model) and to $\tilde{Co}(t+\delta,t+T_r-\delta)$.
	\item $\tilde{CB}(t,t+T_r)$ servers that during a time interval $[t,t+T_r]$ belong to $\tilde{Co}(t+\delta,t+T_r-\delta)$ and to $\tilde{B}(t,t+T_r)$.
	\end{itemize} 
\end{definition}

\begin{definition}[$Sil(t,t+T_r)$]
	Let $[t,t+T_r]$ be a time interval. $Sil(t,t+T_r)$ is the set of servers in $Cu(t,t+T_R-\delta)$. 
\end{definition}

Servers belonging to $Sil(t_B(op_R),t_E(op_R))$ are servers that do no participate to $op_R$. In oder words, those servers in the worst case scenario became correct after $t_E(opR)-\delta$, thus if they send back a correct reply it is not sure that client delivers such reply before the end of $T_r$ time. 
Now we can define the worst case scenarios for the sets we defined so far with respect to $S^*$.

\begin{definition}[$Max\tilde{B}(t,t+T_r)$]
	Let $S$ be a scenario and $[t,t+T_r]$ a time interval.
	The cardinality of $\tilde{B}_S(t,t+T_r)$ is maximum with respect to $S$ if for any $t'$, $t'>0$, we have that $|\tilde{B}_S(t,t+T_r)|\geq |\tilde{B}_S(t',t'+T_r)|$. Then we call the value of such cardinality as $\mathit{Max\tilde{B}_S}(t,t+T_r)$.
	If we consider only one scenario per time then we can omit the subscript related to the scenario and write directly $Max\tilde{B}(t,t+T_r)$.
\end{definition}

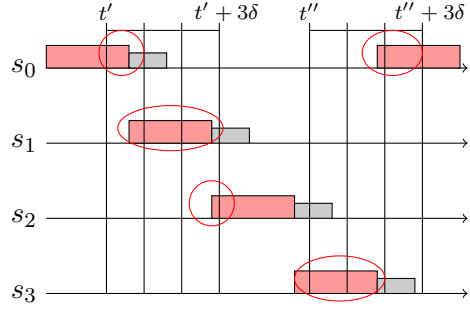
\begin{figure}
		\begin{tikzpicture}[y=-1cm]
		
		\def \lenght {5.6}
		\def \n {3} 
		
		\def \deltaGrande {1.1}
		\def \deltaPiccolo{0.5}
		\def \gammaCuring{0.5}
		
		\processes
		\faults{0}{0}{4}{gray}{0}

		\def \tuno {.8}
		\intervallo{\tuno}{3}{t'}
		
\draw[red] (1,-.2) circle (.3cm);
\draw[red] (1.65,.8) ellipse (.7cm and .3cm);
\draw[red] (2.2,1.8) circle (.3cm);

		\intervallo{\tuno+2.7}{3}{t''}
\draw[red] (4.6,-.2) ellipse (.4cm and .3cm);	
\draw[red] (3.9,2.8) ellipse (.6cm and .3cm);		
		\end{tikzpicture}
		\caption{Let $[t,t+T_r]$ be time a interval such that in the given scenario $|\tilde{B}(t,t+T_r)|=Max\tilde{B}(t,t+T_r)$. In particular we have that in the time interval $[t',t'+T_r]$, $|\tilde{B}(t',t'+T_r)|=Max\tilde{B}(t,t+T_r)$. While in the time interval $[t'',t''+T_r]$, $|\tilde{B}(t'',t''+T_r)|<Max\tilde{B}(t,t+T_r)$. }\label{fig:maxB}
\end{figure}

This value quantifies in the worst case scenario how many servers can be Byzantine, for at least one time unit,  during a \op{read} operation. Figure \ref{fig:maxB} depicts a scenario where $T_r=3\delta$ and during the time interval $[t',t'+T_r]$ there is a maximum number of Byzantine servers while in $[t'',t''+T_r]$ this number is not maximal.

\begin{definition}[$MaxSil(t,t+T_r)$]
	Let $S$ be a scenario and $[t,t+T_r]$ a time interval.
	The cardinality of $Sil_S(t,t+T_r)$ is maximum with respect to $S$ if for any $t'$, $t'\geq 0$ we have that $|Sil(t,t+T_r)|\geq |Sil(t',t'+T_r)|$ and $\tilde{B}(t,t+T_r)=Max\tilde{B}(t,t+T_r)$. Then we call the value of such cardinality as $MaxSil_S(t,t+T_r)$.
	If we consider only one scenario per time then we can omit the subscript related to the scenario and write directly $minSil(t,t+T_r)$. 
\end{definition}

This value quantifies the maximum number of servers that begin in a cured state a \op{read} operation and are still cured after $T_r-\delta$ time. So that any correct reply sent after such period has no guarantees to be delivered by the client and such servers are assumed to be silent.

\begin{definition}[$Max{Cu}(t)$]
	Let $S$ be a scenario and  $t$ be a time instant. The cardinality of $Cu_S(t)$ is maximum with respect to $S$ if for any $t'$, $t'\geq 0$, we have that $|Cu_S(t')| \leq |Cu_S(t)|$ and $\tilde{B}(t,t+T_r)=Max\tilde{B}(t,t+T_r)$. We call the value of such cardinality as $Max{Cu}_S(t)$.
	If we consider only one scenario per time then we can omit the subscript related to it and write directly $Max{Cu}(t)$. 
\end{definition}

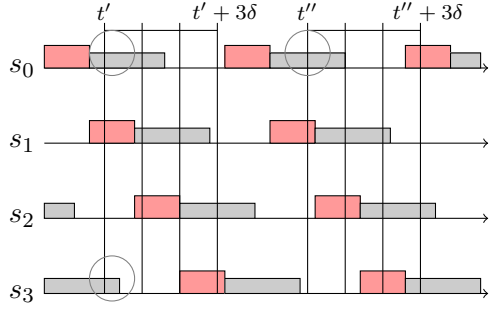
\begin{figure}
		\begin{tikzpicture}[y=-1cm]
		
		\def \lenght {5.9}
		\def \n {3} 
		
		\def \deltaGrande {.6}
		\def \deltaPiccolo{0.5}
		\def \gammaCuring{1}
		
		\processes
		\faults{0}{0}{8}{gray}{1}
		\curatoParziale{9*\deltaGrande}{0}{\lenght-9*\deltaGrande-.1}{gray}

		\def \tuno {.8}
		\intervallo{\tuno}{3}{t'}
		
		\intervallo{\tuno+2.7}{3}{t''}	
		\draw[gray] (3.5,-.2) circle (.3cm);
		
		\draw[gray] (.9,-.2) circle (.3cm);	
		\draw[gray] (.9,2.8) circle (.3cm);		
		\end{tikzpicture}
		\caption{Let us consider the time instant $t$ and the depicted scenario such that $|Cu(t)|=MaxCu(t)$. In particular, in this case $|Cu(t')|=MaxCu(t)$ and $|Cu(t'')|<MaxCu(t)$.  }\label{fig:maxCu}
\end{figure}

This value quantifies, in the worst case scenario, how many cured servers there may be at the beginning of a \op{read} operation. Figure \ref{fig:maxCu} depicts a scenario where at time $t'$ there are the maximum number of cured server while at $t''$ this value is not maximum. Notice that in such figure, in case of a shorter time interval $[t',t'+2\delta]$ $s_0$ would be silent.

\begin{definition}[$min\tilde{Co}(t,t+T_r)$] 
	Let $S$ be a scenario and $[t,t+T_r]$ be a time interval then $min\tilde{C}_S(t,t+T_r)$ denotes the minimum number of correct servers during a time interval $[t+\delta,t+T_r-\delta]$. If we consider only one scenario per time then we can omit the subscript related to it and write directly $min\tilde{C}(t,t+T_r)$.
\end{definition}

\begin{figure}
		\begin{tikzpicture}[y=-1cm]
		
		\def \lenght {5.6}
		\def \n {3} 
		
		\def \deltaGrande {1.1}
		\def \deltaPiccolo{0.5}
		\def \gammaCuring{0.5}
		
		\processes
		\faults{0}{0}{4}{gray}{0}

		\def \tuno {.8}
		\intervallo{\tuno}{3}{t'}
\draw[blue] (1.55,-.2) ellipse (.3cm and .3cm);		
\draw[blue] (1.55,1.8) circle (.3cm);
\draw[blue] (1.55,2.8) circle (.3cm);

		\intervallo{\tuno+2.7}{3}{t''}
\draw[blue] (4.25,-.2) ellipse (.3cm and .3cm);
\draw[blue] (4.25,.8) ellipse (.3cm and .3cm);	
\draw[blue] (4.25,1.8) ellipse (.3cm and .3cm);		
		\end{tikzpicture}
		\caption{Let $[t,t+T_r]$ be a time interval such that in the depicted scenario $|\tilde{Co}(t,t+T_r)|=min\tilde{Co}(t,t+T_r)$. Then in both time intervals $[t',t'+T_r]$ and $[t'',t''+T_r]$ we have that $|\tilde{Co}(t',t'+T_r)|=|\tilde{Co}(t'',t''+T_r)|=min\tilde{Co}(t,t+T_r)$.}\label{fig:minCo}
\end{figure}
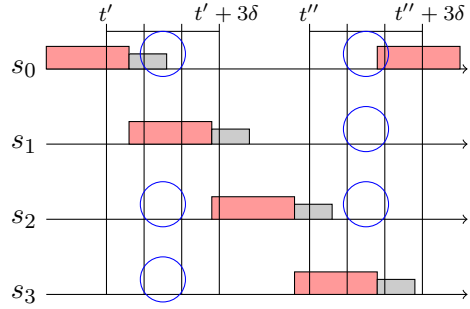

Figure \ref{fig:minCo} depicts a scenario where during the both  intervals $[t',t'+T_r]$ and $[t'',t''+T_r]$ the  number of correct servers is minimum.

\begin{definition}[$min\tilde{CBC}(t,t+T_r)$]
	Let $[t,t+T_r]$ be a time interval then $min\tilde{CBC}(t,t+T_r)$ denotes the minimum number of servers that during a time interval $[t,t+T_r]$ belong first to $\tilde{B}(t,t+T_r)$ or $Cu(t)$ (only in $(\Delta S, CUM)$ model) and then to $Co(t+\delta,t+T_r-\delta)$ or vice versa and $\tilde{B}(t,t+T_r)=Max\tilde{B}(t,t+T_r)$.\\
	In particular let us denote as:
	\begin{itemize}
	\item $min\tilde{BC}(t,t+T_r)$ the minimum number of servers that during a time interval $[t,t+T_r]$ belong to $\tilde{B}(t,t+T_r)$ or $Cu(t)$ (only in $(\Delta S, CUM)$ model) and to $\tilde{Co}(t+\delta,t+T_r-\delta)$.
	\item $min\tilde{CB}(t,t+T_r)$ the minimum number of servers that during a time interval $[t,t+T_r]$ belong to $\tilde{Co}(t+\delta,t+T_r-\delta)$ and to $\tilde{B}(t,t+T_r)$.
	\end{itemize} 
\end{definition}

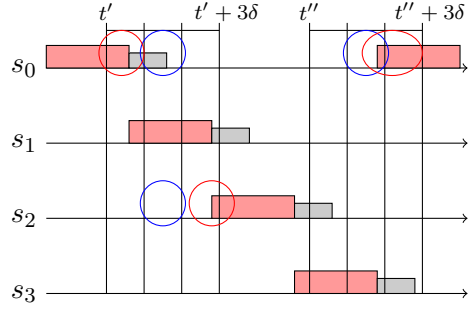
\begin{figure}
		\begin{tikzpicture}[y=-1cm]
		
		\def \lenght {5.6}
		\def \n {3} 
		
		\def \deltaGrande {1.1}
		\def \deltaPiccolo{0.5}
		\def \gammaCuring{0.5}
		
		\processes
		\faults{0}{0}{4}{gray}{0}

		\def \tuno {.8}
		\intervallo{\tuno}{3}{t'}
\draw[blue] (1.55,-.2) circle (.3cm);		
\draw[blue] (1.55,1.8) circle (.3cm);

\draw[red] (1,-.2) circle (.3cm);
\draw[red] (2.2,1.8) circle (.3cm);

		\intervallo{\tuno+2.7}{3}{t''}
\draw[blue] (4.25,-.2) ellipse (.3cm and .3cm);
\draw[red] (4.6,-.2) ellipse (.4cm and .3cm);	

		\end{tikzpicture}
		\caption{Let $[t,t+T_r]$ a time interval such that in the depicted scenario $\tilde{CBC}(t,t+T_r)=min\tilde{CBC}(t,t+T_r)$. Then  $\tilde{CBC}(t',t'+T_r)>min\tilde{CBC}(t,t+T_r)$ and  $\tilde{CBC}(t'',t''+T_r)=min\tilde{CBC}(t,t+T_r)$.}\label{fig:minCBC}
\end{figure}

As we stated before, Byzantine servers set changes during the \op{read} operation $op_R$, so there can be servers that are in a Byzantine state at $t_B(op_R)$ and in a correct state before $t_E(op_R)-\delta$ (cf. $s_0$ during $[t',t'+3\delta]$ time interval in Figure \ref{fig:minCBC}). Those servers contribute with an incorrect message at the beginning and with a correct message after. The same may happen with servers that are correct from $t_B(op_R)$ to at least $t_B(op_R)+\delta$ (so that for sure deliver the read request message and send the reply back) and are affected by a mobile agent after $t_B(op_R)+\delta$ (cf. $s_0$ during $[t'',t''+3\delta]$ time interval in Figure \ref{fig:minCBC}).  

\begin{lemma}\label{l:MaxB}
	$Max\tilde{B}(t,t+T_r)=(\lceil \frac{T_r}{\Delta} \rceil +1)f$. 
\end{lemma}

\begin{proofL}	
	For simplicity let us consider a single agent $ma_k$, then we extend the same reasoning to all the $f$ agents. In $[t,t+T_r]$ time interval, with $T_r \geq 2\delta$, $ma_k$ can affect a different server each $\Delta$ time. It follows that the number of times it may change server is $ \frac{T_r}{\Delta}$. Thus the affected servers are $\lceil \frac{T_r}{\Delta} \rceil$ plus the server that was affected at $t$. Finally, extending the reasoning to $f$ agents, $Max\tilde{B}(t,t+T_r)=(\lceil \frac{T_r}{\Delta} \rceil +1)f$, which concludes the proof.	   
	\renewcommand{\toto}{l:MaxB}
\end{proofL}

As we see in the sequel, the value of $Max\tilde{B}(t,t+T_r)$ is enough to compute the lower bound. Now we can define the worst case scenario for a \op{read} operation with respect to $S^*$. Let $op$ be a ${\sf read}$ operation issued by $c_i$.
We want to define, among the messages that can be deliver by $c_i$ during $op$, the minimum amount of messages sent by server when they are in a correct state and the maximum amount of messages sent by servers when they are not in a correct state.\\
In each scenario, we assume that each message sent to or by Byzantine servers is instantaneously delivered, while each message sent to or by correct servers requires $\delta$ time. Without loss of generality, let us assume that all Byzantine servers send the same value and send it only once, for each period where they are Byzantine. Moreover, we make the assumption that each cured server (in the CAM model) does not reply as long as it is cured. Yet, in the CUM model, it behaves similarly to Byzantine servers, with the same assumptions on message delivery time.

\begin{definition}[$MaxReplies\_NCo(t,t+T_r)_k$]
Let $MaxReplies\_NCo(t,t+T_r)_k$ be the multi-set maintained by client $c_k$ containing $m_{ij}$ elements, where $m_{ij}$ is the $i-th$ message delivered by $c_k$ and sent at time $t', t' \in[t,t+T_r]$ by $s_j$ such that $s_j \notin Co(t')$.    
\end{definition}

Considering the definitions of both $Max\tilde{B}(t,t+T_r)$ and $MaxCu(t)$ the next Corollary follows:

\begin{corollary}\label{c:maximum}
In the worst case scenario, during a ${\sf read}$ operation lasting $T_r \geq 2\delta$ issued by client $c_i$, $c_i$ delivers $Max\tilde{B}(t,t+T_r)$ incorrect replies in the $(\Delta S, CAM)$ model and $Max\tilde{B}(t,t+T_r)+MaxCu(t)$ incorrect replies in the $(\Delta S, CUM)$ model . 
\end{corollary}

\begin{definition}[$minReplies\_Co(t,t+T_r)_k$]
Let $minReplies\_Co(t,t+T_r)_k$ be the multi-set maintained by client $c_k$ containing $m_{ij}$ elements, where $m_{ij}$ is the $i-th$ message delivered by $c_k$ and sent at time $t', t' \in[t,t+T_r]$ by $s_j$ such that $s_j \in Co(t')$.  
\end{definition}

Note that correct replies come from servers that \emph{(i)} have never been affected during the time interval $[t,t+T_r]$, or \emph{(ii)} where in a cured state at $t$ but do not belong to the $Sil(t,t+T_r)$ set, or \emph{(iii)} servers that reply both correctly and incorrectly. The next Corollary follows.

\begin{corollary}\label{c:minimum}
In the worst case scenario, during a ${\sf read}$ operation lasting $T_r \geq 2\delta$ issued by client $c_i$, $c_i$ delivers $n-(Max\tilde{B}(t,t+T_r)+MaxSil(t,t+T_r))+ min\tilde{CBC}(t,t+T_r)$ correct replies in the $(\Delta S, CAM)$ model and $n-[Max\tilde{B}(t,t+T_r)+ MaxCu(t)] + min\tilde{CBC}(t,t+T_r)$ correct replies in the $(\Delta S, CUM)$ model. 
\end{corollary}

In the following, given a time interval, we characterize correct and incorrect servers involved in such interval. Concerning correct servers, let us first analyze when a client collects  $x \leq n$ different replies and then we extend such result to  $x > n$. Then we do the same for incorrect replies.

\begin{lemma}\label{l:CorSetComposition}
Let $op$ be a ${\sf read}$ operation issued by client $c_i$ in a scenario $S^*$, whose duration is $T_r \geq 2\delta$. Let $x, x \geq 2$, be the number of messages delivered by $c_i$ during $op$. If $x \leq n$ then $minReplies\_Co(t,t+T_r)_k$ contains replies from $x$ different servers.  
\end{lemma}

\begin{proofL}
Let us suppose that $minReplies\_Co(t,t+T_r)_k$ contains replies from $x-1$ different servers (trivially it can not be greater than $x$). Without lost of generality, let us suppose that $c_i$ collects replies from $s_1, \dots, s_{x-1}$. It follows that there is a server $s_i, i \in [1, x-1]$ that replied twice and a server $s_x$ that did not replied. Let us also suppose \emph{w.l.g.} that there is one Byzantine mobile agent $ma_k$ (i.e., $f=1$).
If during the time interval $[t,t+T_r]$ $s_x$ never replied, then $s_x$ has been affected at least during $[t+\delta, t+T_r-\delta - \gamma +1]$. This implies that $T_r\leq \Delta +2\delta + \gamma$. Since $s_i$ replies twice then two scenarios are possible during $op$: \emph{(i)} $s_i$ was first affected by $ma_k$ and then became correct (so it replied once), then affected again and then correct again (so it replied twice); \emph{(ii)} $s_i$ was correct (so it replied once), then it was affected by $ma_k$ and then correct again (so it replied twice). Let us consider case \emph{(ii)} (case \emph{(i)} follows trivially). Since $s_i$ had the time to reply ($\delta$), to be affected and then became correct ($\Delta+\gamma$) and reply again ($\delta$) this means that $T_r>\Delta+2\delta+\gamma$. A similar result we get in case \emph{(i)} where the considered execution requires a longer time. This is in contradiction with $T_r\leq \Delta +2\delta + \gamma$ thus $c_i$ gets replies for $x$ different servers. 
\renewcommand{\toto}{l:CorSetComposition}
\end{proofL}


If a client delivers $n>x$ messages then we can apply the same reasoning of the previous Lemma to the first chunk of $n$ messages, then to the second chunk of $n$ messages and so on. Roughly speaking, if $n=5$ and a client delivers $11$ messages from correct processes, then there are $3$ occurrences of the message coming from the first server and $2$ occurrences of the messages coming from the remaining servers. Thus the next Corollary directly follows.

\begin{corollary}\label{c:CorSetComposition}
Let $op$ be a ${\sf read}$ operation issued by client $c_i$ in a scenario $S^*$, $op$ duration is $T_r \geq 2\delta$. Let $x, x \geq 2$, be the number of messages delivered by $c_i$ during $op$, then $minReplies\_Co(t,t+T_r)_k$ contains $x \mod{n}$ messages $m_{ij}$ whose occurrences is $\lfloor \frac{x}{n} \rfloor +1$ and  $(n - x \Mod{n})$ messages whose occurrences is $\lfloor \frac{x}{n} \rfloor$.
\end{corollary}

The case of $MaxReplies\_NCo(t,t+T_r)_k$ directly follows from scenario $S^*$, since by hypotheses mobile Byzantine agents move circularly from servers to servers, never passing on the same server before having affected all the others. Thus,  the following corollary holds. 

\begin{corollary}\label{c:ByzSetComposition}
Let $op$ be a ${\sf read}$ operation issued by client $c_i$ in a scenario $S^*$, $op$ duration is $T_r \geq 2\delta$. Let $x, x \geq 2$, be the number of messages delivered by $c_i$ during $op$, then $MaxReplies\_NCo(t,t+T_r)_k$ contains $x \mod{n}$ messages $m_{ij}$ whose occurrences is $\lfloor \frac{x}{n} \rfloor +1$ and  $(n - x \Mod{n})$ messages whose occurrences is $\lfloor \frac{x}{n} \rfloor$.
\end{corollary}

At this point we can compute how many correct and incorrect replies a client $c_k$ can deliver in the worst case scenario during a time interval $[t,t+T_r]$. Trivially, $c_k$ in order to distinguish correct and incorrect replies needs to get $minReplies\_Co(t,t+T_r)_k > MaxReplies\_NCo(t,t+T_r)_k$. It follows that the number of correct servers has to be enough to guarantee this condition. Table \ref{tab:summary} follows directly from this observation. In a model with $b$ Byzantine (non mobile) a client $c_i$ requires to get at least $2b+1$ replies to break the symmetry and thus $n\geq 2b+1$. In presence of mobile Byzantine we have to sum also servers that do not reply (silent) and do not count twice servers that reply with both incorrect and correct values.

\begin{table*}[t]
\centering
	\scriptsize
	\begin{tabular}{| c | c |}
		\hline
		 $n_{CAM_{LB}}$ & $[2Max\tilde{B}(t,t+T_r)+MaxSil(t,t+T_r)-min\tilde{CBC}(t,t+T_r)]f $ \\
		\hline
		 $n_{CUM_{LB}}$ & $[2(Max\tilde{B}(t,t+T_r)+MaxCu(t,t+T_r))-min\tilde{CBC}(t,t+T_r)]f  $ \\
		\hline		
	\end{tabular}
	\caption{Lower bounds on the number of replicas in each model.}\label{tab:summary}
\end{table*}

\begin{theorem}\label{t:LB}
	If $n < n_{CAM_{LB}}$ ($n < n_{CUM_{LB}}$) as defined in Table \ref{tab:summary}, then there not exists a protocol $\mathcal{P}_{reg}$ solving the safe register specification in $(\Delta S,CAM)$ model ($(\Delta S, CUM)$ model respectively).
\end{theorem}

\begin{proofT}
	Let us suppose that $n < n_{CAM_{LB}}$ ($n < n_{CUM_{LB}}$) and that protocol $\mathcal{P}_{reg}$ does exist. If a client $c_i$ invokes a ${\sf read}$ operation $op$, lasting $T_r \geq 2 \delta$ time, if no ${\sf write}$ operations occur, then $c_i$ returns a valid value at time $t_B(op)$. 
	Let us consider an execution $E_0$ where $c_i$ invokes a ${\sf read}$ operation $op$ and let $0$ be the valid value at $t_B(op)$. Let us assume that all Byzantine severs involved in such operation reply once with $1$. From Corollaries \ref{c:maximum} and \ref{c:minimum}, $c_i$ collects $MaxReplies\_NCo(t,t+T_r)_i$ occurrences of $1$ and $minReplies\_Co(t,t+T_r)_i$ occurrences of $0$. Since $\mathcal{P}_{reg}$ exists and no ${\sf write}$ operations occur, then $c_i$ returns $0$. Let us now consider a another execution $E_1$ where $c_i$ invokes a ${\sf read}$ operation $op$ and let $1$ be the valid value at $t_B(op)$. Let us assume that all Byzantine severs involved in such operation replies once with $0$. From Corollaries \ref{c:maximum} and \ref{c:minimum} and Corollary \ref{c:CorSetComposition} and Corollary \ref{c:ByzSetComposition}, $c_i$ collects $MaxReplies\_NCo(t,t+T_r)_i$ occurrences of $0$ and $minReplies\_Co(t,t+T_r)_i$ occurrences of $1$. Since $\mathcal{P}_{reg}$ exists and no ${\sf write}$ operations occur, then $c_i$ returns $1$.

	From Lemma \ref{l:MaxB} and using values in Table \ref{tab:summary} we obtain following equations for both models:
	\begin{itemize}
	\item{$(\Delta S, CAM)$:}
		\begin{itemize}
		\item{$MaxReplies\_NCo(t,t+T_r)_i$=} $Max\tilde{B}(t,t+T_r) = (\lceil \frac{T_r}{\Delta} \rceil +1)f$
		\item{$minReplies\_Co(t,t+T_r)_i$=} $n-[Max\tilde{B}(t,t+T_r)+MaxSil(t,t+T_r)]+ min\tilde{CBC}(t,t+T_r) =$
		\end{itemize}
		$$[2(Max\tilde{B}(t,t+T_r))+MaxSil(t,t+T_r)$$
                $$-min\tilde{CBC}(t,t+T_r)]$$
                $$ - [(Max\tilde{B}(t,t+T_r)+MaxSil(t,t+T_r))$$
                $$ + min\tilde{CBC}(t,t+T_r)]=$$
		$$Max\tilde{B}(t,t+T_r) =(\lceil \frac{T_r}{\Delta} \rceil +1 )f$$

	\item{$(\Delta S, CUM)$:}
		\begin{itemize}
		\item{$MaxReplies\_NCo(t,t+T_r)_i$=} $Max\tilde{B}(t,t+T_r) + MaxCu(t) =
		(\lceil \frac{T_r}{\Delta} \rceil +1  )f+ MaxCu(t)$
		\item{$minReplies\_Co(t,t+T_r)_i$=} $n-[Max\tilde{B}(t,t+T_r) + MaxCu(t)] + min\tilde{CBC}(t,t+T_r)=$ 
			\end{itemize}
		$$[2Max\tilde{B}(t,t+T_r)+2MaxCu(t))-min\tilde{CBC}(t,t+T_r)]+$$ $$ - [Max\tilde{B}(t,t+T_r) + MaxCu(t)] + min\tilde{CBC}(t,t+T_r)= $$
		$$Max\tilde{B}(t,t+T_r)+MaxCu(t)= (\lceil \frac{T_r}{\Delta} \rceil +1 )f + MaxCu(t)$$

	\end{itemize}	

 It follows that in $E_0$ and $E_1$ $c_i$ delivers the same occurrences of $0$ and $1$, both executions are indistinguishable leading to a contradiction.
 	
	\renewcommand{\toto}{t:LB}
\end{proofT}

$MaxReplies\_NCo(t,t+T_r)_i$ and $minReplies\_Co(t,t+T_r)_i$ are equal independently from the value assumed by $T_r$, the \op{read} operation duration. From the equation just used in the previous lemma the next Corollary follows.

\begin{corollary}\label{cor:maCheMeFregaDiT}
For each $T_r \geq 2\delta$ if $n > n_{CAM_{LB}}$ ($n > n_{CUM_{LB}}$) then $MaxReplies\_NCo(t,t+T_r)_i < minReplies\_Co(t,t+T_r)_i$.
\end{corollary}

At this point we compute $minCu(t)$, $MaxSil(t,t+T_r)$ and $min\tilde{CBC}(t,t+T_r)$ to finally state exact lower bounds depending on the system parameters, in particular depending on $\Delta$, $\gamma$ and the servers awareness, i.e., $(\Delta S,CAM)$ and $(\Delta S,CUM)$. 


Let us adopt the following notation. Given the time interval $[t,t+T_r]$ let $\{s_1,s_2,\dots,s_b\} \in B(t,t+T_r)$ be the servers affected sequentially during $T_r$ by the mobile agent $ma_k$. Let $\{s_{-1}, s_{-2}, \dots , s_{-c}\} \in Cu(t)$ be the servers in a cured state at time $t$ such that $s_{-1}$ is the last server that entered in such state and $s_{c}$ the first server that became cured. Let  $t_BB(s_i)$ and $t_EB(s_i)$ be respectively the time instant in which $s_i$ become Byzantine and the time in which the Byzantine agent left. $t_BCu(s_i)$ and $t_ECu(s_i)$ are respectively the time instant in which $s_i$ become cured and the time instant in which it became correct. 
Considering that $ma_k$ moves each $\Delta$ time then we have that $t_BB(s_{i-1})-t_BB(s_{i})=\Delta$ and $t_BCu(s_{-j})-t_BCu(s_{-j+1})=\Delta$. The same holds for the $t_E$ of such states. Moreover $t_BB(s_1)=t_BCu(s_{-1})$.
Now we are ready to build the read scenario with respect to $S^*$. In particular we build a scenario for the $(\Delta S, CAM)$ model and one for the $(\Delta S, CUM)$ model. Intuitively, the presence of cured servers do not have the same impact in the two models, thus in the $(\Delta S, CUM)$ model we maximize such number.
Let $[t,t+2\delta]$ be the considered time interval and let $\epsilon$ be a positive number arbitrarily smaller, then we consider in the $(\Delta S,CAM)$ scenarios $t=t_EB(s_1)-\epsilon$ (cf. Figure \ref{fig:scenarioConLetturaCAM}) and in the $(\Delta S,CUM)$ scenarios $t_BB(s_b)=t+2\delta-\epsilon$ (cf. Figure \ref{fig:scenarioConLetturaCUM}).

\begin{figure}
\centering
		\begin{tikzpicture} [y=-1cm]
		
		\def \lenght {5.3}
		\def \n {3} 
		
		\def \deltaGrande {0.6}
		\def \deltaPiccolo{0.5}
		\def \gammaCuring{1}
		

		\foreach \x in {0,...,\n} 
		\draw[->] (0,\x) -- (\lenght,\x); 
\node[]() at (0,-.4) {\scriptsize{$s_{-2=-c}$}}; 
\node[]() at (.6,.6) {\scriptsize{$s_{1}$}}; 
\node[]() at (1.2,1.6) {\scriptsize{$s_{2}$}}; 
\node[]() at (0,2.6) {\scriptsize{$s_{-1}$}};
\node[]() at (2.2,2.6) {\scriptsize{$s_{3=b}$}};

		
		\node[]() at (0+8*\deltaGrande, -.6) {\scriptsize{$\dots$}};
		

		\faults{0}{0}{7}{green}{1}
		\curatoParziale{8*\deltaGrande}{3}{\lenght-8*\deltaGrande-.1}{green}

		\intervallo{1.1}{2}{t}		
		\end{tikzpicture}
		\caption{Representation of  $S^*$ when we consider a $(\Delta S,CAM)$ model, in particular $t_EB(s_1)=t+\epsilon$, for $\epsilon>0$ and arbitrarily small.}\label{fig:scenarioConLetturaCAM}
\end{figure} 

\begin{figure}
\centering
		\begin{tikzpicture} [y=-1cm]
		
		\def \lenght {5.3}
		\def \n {3} 
		
		\def \deltaGrande {0.6}
		\def \deltaPiccolo{0.5}
		\def \gammaCuring{1}
		

		\foreach \x in {0,...,\n} 
		\draw[->] (0,\x) -- (\lenght,\x); 
\node[]() at (0,-.4) {\scriptsize{$s_{-2=-c}$}}; 
\node[]() at (.6,.6) {\scriptsize{$s_{1}$}}; 
\node[]() at (1.2,1.6) {\scriptsize{$s_{2}$}}; 
\node[]() at (0,2.6) {\scriptsize{$s_{-1}$}};
\node[]() at (2.2,2.6) {\scriptsize{$s_{3=b}$}};

		
		\node[]() at (0+8*\deltaGrande, -.6) {\scriptsize{$\dots$}};
		

		\faults{0}{0}{7}{yellow}{1}
		\curatoParziale{8*\deltaGrande}{3}{\lenght-8*\deltaGrande-.1}{yellow}

		\intervallo{.9}{2}{t}		
		\end{tikzpicture}
		\caption{Representation of  $S^*$ when we consider a $(\Delta S,CUM)$ model, in particular $t_BB(s_c)=t+2\delta-\epsilon$, for $\epsilon>0$ and arbitrarily small.}\label{fig:scenarioConLetturaCUM}
\end{figure} 


In the sequel we use the notion of Ramp Function:
 $$\mathcal{R}(x) = \begin{cases} x &\mbox{if } x \geq 0 \\
0 & \mbox{if } x < 0 \end{cases}  $$

\begin{lemma}\label{lem:maxCuCAM}
Let us consider a time interval $[t,t+T_r], T_r\geq 2\delta$ and an arbitrarily small number $\epsilon>0$, then in fthe $(\Delta S,CAM)$ model $MaxCu(t)= \mathcal{R}(\ceil{\frac{\gamma-\Delta +\epsilon}{\Delta}})$.
\end{lemma}

\begin{proofL}
As we defined, $s_{-1}$ is the most recent server that entered in a cured state, with respect to the considered time interval. Intuitively each $s_{-j}$ is in $Cu(t)$ if $t_ECu(s_{-j})>t$. Considering that $t_ECu(s_{-j})-t_ECu(s_{-j-1})=\Delta$ then the number of servers in a cured state at $t$ is $MaxCu(t)= \ceil{\frac{t_ECu(s_1)-t}{\Delta}}$. \footnote{Consider Figure \ref{fig:scenarioConLetturaCAM}, $s_2$ is the most recent server that entered in the cured state. This is the server that spend more time in such state with respect to the others. It follows that other servers are in a cured state if during this time interval there is enough time for a \vir{jump}} As we stated, for $(*,CAM)$ models we consider scenarios in which $t$, the beginning of the considered time interval, is just before $t_EB(s_1)$. Thus given an arbitrarily small number $\epsilon>0$, let $t=t_EB(s_1)-\epsilon$. By construction we know that $t_BB(s_1)=t_EB(s_1)-\Delta=t_BCu(s_{-1})$. Substituting $t_BCu(s_{-1})=t+\epsilon-\Delta$, since we consider $\gamma$ the upper bound for the curing time, then $t_ECu(s_{-1})=t+\epsilon-\Delta+\gamma$ .  So finally, $MaxCu(t)= \ceil{\frac{t_ECu(s_1)-t}{\Delta}} = \ceil{\frac{\gamma-\Delta +\epsilon}{\Delta}}$ and since there can no be a negative result then $MaxCu(t)= \mathcal{R}(\ceil{\frac{\gamma-\Delta +\epsilon}{\Delta}})$. This concludes the proof.  
\renewcommand{\toto}{lem:maxCuCAM}
\end{proofL}

\begin{lemma}\label{lem:maxCuCUMGenerale}
Let us consider a time interval $[t,t+T_r], T_r\geq 2\delta$ and an arbitrarily small number $\epsilon>0$, then in the $(\Delta S,CUM)$ model $MaxCu(t) = \mathcal{R}( \ceil{\frac{T_r - \epsilon - \ceil{\frac{T_r}{\Delta}}\Delta +\gamma}{\Delta}})$.
\end{lemma}

\begin{proofL}
As we defined, $s_{-1}$ is the most recent server that entered in a cured state, with respect to the considered interval. Intuitively, $s_{-j}$ is in $Cu(t)$ if $t_ECu(s_{-j})>t$. Considering that $t_ECu(s_{-j})-t_ECu(s_{-j-1})=\Delta$ then the number of servers in a cured state at $t$ is $MaxCu(t)= \ceil{\frac{t_ECu(s_1)-t}{\Delta}}$. As we state, for $(*,CUM)$ models we consider scenarios in which the end of the considered time interval, is just after $t_BB(s_b)$. Thus given an arbitrarily small number $\epsilon>0$, let $t_BB(s_b)=t+T_r-\epsilon$. By construction we know that $t_BB(s_1)=t_EB(s_1)-\Delta=t_BCu(s_{-1})$ and $t_BB(s_1)=t_BB(s_b)-\ceil{\frac{T_r}{\Delta}}\Delta$ (cf. Lemma \ref{l:MaxB}). Substituting and considering that $t_ECu(s_{-1})=t_BCu(s_{-1})+\gamma)$ we get the following: $t_ECu(s_{-1}) = t + T_r - \epsilon - \ceil{\frac{T_r}{\Delta}}+\gamma$. Finally $MaxCu(t)= \ceil{\frac{t_ECu(s_1)-t}{\Delta}} = \ceil{\frac{T_r - \epsilon - \ceil{\frac{T_r}{\Delta}}+\gamma}{\Delta}}$ and since there can not be a negative result then $MaxCu(t)= \mathcal{R}( \ceil{\frac{T_r - \epsilon - \ceil{\frac{T_r}{\Delta}}\Delta +\gamma}{\Delta}})$. This concludes the proof.
\renewcommand{\toto}{lem:maxCuCUMGenerale}
\end{proofL}

\begin{lemma}\label{lem:maxSilCAM}
Let us consider a time interval $[t,t+T_r], T_r\geq 2\delta$ and an arbitrarily small number $\epsilon>0$, then in the $(\Delta S,CAM)$ model $MaxSil(t,t+T_r)=  \mathcal{R}(\ceil{\frac{\gamma-\Delta +\epsilon-T_r+\delta}{\Delta}})$.
\end{lemma}

\begin{proofL}
As we defined, $s_{-1}$ is the most recent server that entered in a cured state, with respect to the considered interval. Intuitively, $s_{-j}$ is in $Sil(t,t+2\delta)$ if $t_ECu(s_{-j})>T_r-\delta$. Considering that $t_ECu(s_{-j})-t_ECu(s_{-j-1})=\Delta$ then the number of servers in a silent state at $t$ is $MaxSil(t,t+2\delta)= \ceil{\frac{t_ECu(s_1)-T_r+\delta}{\Delta}}$. As we stated for $(\Delta S,CAM)$ models we consider scenarios in which $t$, the beginning of the considered time interval, is just before $t_EB(s_1)$. Thus given an arbitrarily small number $\epsilon>0$, let $t=t_EB(s_1)-\epsilon$. By construction we know that $t_BB(s_1)=t_EB(s_1)-\Delta=t_BCu(s_{-1})$. Substituting $t_BCu(s_{-1})=t+\epsilon-\Delta$, since we consider $\gamma$ the upper bound for curing time, then $t_ECu(s_{-1})=t+\epsilon-\Delta+\gamma$ . So finally, $MaxSil(t,t+T_r)= \ceil{\frac{t_ECu(s_1)-T_r+\delta}{\Delta}} = \ceil{\frac{\gamma-\Delta +\epsilon -T_r+\delta}{\Delta}}$, then since there can not be a negative result $MaxSil(t,t+2\delta)=  \mathcal{R}(\ceil{\frac{\gamma-\Delta +\epsilon-T_r+\delta}{\Delta}})$.
\renewcommand{\toto}{lem:maxSilCAM}
\end{proofL}

\begin{lemma}\label{lem:maxSilCUMGenerale}
Let us consider a time interval $[t,t+T_r], T_r\geq 2\delta$ and an arbitrarily small number $\epsilon>0$, then in the $(\Delta S,CUM)$ model $MaxSil(t,t+T_r)= \ceil{\frac{T_r - \epsilon - \ceil{\frac{T_r}{\Delta}}\Delta +\gamma-\delta}{\Delta}}$.
\end{lemma}

\begin{proofL}
As we defined, $s_{-1}$ is the most recent server that entered in a cured state, with respect to the considered interval. Intuitively, $s_{-j}$ is in $Sil(t,t+T_r)$ if $t_ECu(s_{-j})>T_r-\delta$. Considering that $t_ECu(s_{-j})-t_ECu(s_{-j-1})=\Delta$ then the number of servers in a silent state at $t$ is $MaxSil(t,t+T_r)= \ceil{\frac{t_ECu(s_1)-T_r+\delta}{\Delta}}$. As we stated for $(\Delta S,CUM)$ models we consider scenarios in which $t+T_r$, the end of the considered time interval, is just after $t_BB(s_b)$. Thus given an arbitrarily small number $\epsilon>0$, let $t_BB(s_b)=t+T_r-\epsilon$. By construction we know that $t_BB(s_1)=t_EB(s_1)-\Delta=t_BCu(s_{-1})$ and $t_BB(s_1)=t_BB(s_b)-\ceil{\frac{T_r}{\Delta}}\Delta$ (cf. Lemma \ref{l:MaxB}). Substituting and considering that $t_ECu(s_{-1}=t_BCu(s_{-1})+\gamma)$ we get the following: $t_ECu(s_{-1} = t + T_r - \epsilon - \ceil{\frac{T_r}{\Delta}}+\gamma$. Finally $MaxSil(t,t+T_r)= \ceil{\frac{t_ECu(s_1)-T_r+\delta}{\Delta}} = \ceil{\frac{T_r - \epsilon - \ceil{\frac{T_r}{\Delta}}+\gamma-T_r+\delta}{\Delta}}$, then since there can not be a negative result, $MaxSil(t,t+T_r)= \ceil{\frac{T_r - \epsilon - \ceil{\frac{T_r}{\Delta}}\Delta +\gamma-T_r+\delta}{\Delta}}$.
\renewcommand{\toto}{lem:maxSilCUMGenerale}
\end{proofL}

\begin{lemma}\label{l:minCBCCAMGenerale}
Let us consider a time interval $[t,t+T_r], T_r\geq 2\delta$ then in the $(\Delta S,CAM)$ model. $min\tilde{CBC}=\mathcal{R}(\ceil{\frac{T_r}{\Delta}}-\ceil{\frac{\delta}{\Delta}})+\mathcal{R}(\ceil{\frac{T_r-\gamma-T_r+\delta}{\Delta}})$. 
\end{lemma}

\begin{proofL}
By definition $min\tilde{CBC}(t,t+T_r)=min\tilde{CB}(t,t+T_r)+min\tilde{BC}(t,t+T_r)$.\\
\noindent {\bf -} $min\tilde{CB}(t,t+T_r)$ is the minimum number of servers that correctly reply and then, before $t+T_r$ are affected and incorrectly reply. 
Let us observe that a correct server correctly reply if belongs to $Co(t,t+\delta)$, it follows that servers in $\tilde{B}(t,t+\delta)$ do not correctly reply.
Thus, $min\tilde{CB}(t,t+T_r)=Max\tilde{B}(t,t+T_r)-Max\tilde{B}(t,t+\delta)$. It may happen that $Max\tilde{B}(t,t+T_r)<Max\tilde{B}(t,t+T_r-\delta)$, but obviously there can no be negative servers, so we consider only non negative values,  $min\tilde{CB}(t,t+T_r)=\mathcal{R}(Max\tilde{B}(t,t+T_r)-Max\tilde{B}(t,t+\delta))$.\\

\noindent {\bf -} $min\tilde{BC}(t,t+2\delta)$ is the minimum number of servers that incorrectly reply and then become correct in time that the correct reply is delivered. A server is able to correctly reply if it is correct before $t+T_r-\delta$ (the reply message needs at most $\delta$ time to be delivered). Thus we are interested in servers that are affected by a mobile agent up to $t+T_r-\gamma-\delta$. For $(\Delta,CAM)$ models we consider scenarios in which $t$, the beginning of the considered time interval, is just before $t_EB(s_1)$. Thus given an arbitrarily small number $\epsilon>0$, let $t=t_EB(s_1)-\epsilon$. In the time interval $[t,t+T_r-\gamma-\delta]$ the number of the mobile agent \vir{jumps} is given by $\ceil{\frac{T_r-\gamma-\delta}{\Delta}}$ 
Trivially, we can not have a negative number, so it becomes $\mathcal{R}(\ceil{\frac{T_r-\gamma-\delta}{\Delta}})$.
Summing up $min\tilde{CBC}=\mathcal{R}(\ceil{\frac{T_r}{\Delta}}-\ceil{\frac{\delta}{\Delta}})+\mathcal{R}(\ceil{\frac{T_r-\gamma-\delta}{\Delta}})$, which concludes the proof.
\renewcommand{\toto}{l:minCBCCAMGenerale}
\end{proofL}

\begin{lemma}\label{l:minCBCUMGenerale}
Let us consider a time interval $[t,t+T_r], T_r\geq 2\delta$, let $\epsilon>0$ be an arbitrarily small number. If $maxCu(t)>0$ or $\gamma>\Delta$ then in the $(\Delta S,CUM)$ model $min\tilde{CB}=\ceil{\frac{T_r-\epsilon-\delta}{\Delta}}$ otherwise $min\tilde{CB}=\mathcal{R}(Max\tilde{B}(t,t+T_r)-Max\tilde{B}(t,t+T_r-\delta))$.
\end{lemma}

\begin{proofL}
$min\tilde{CB}(t,t+T_r)$ is the minimum number of servers that correctly reply and then, before $t+T_r$ are affected by a mobile agent and incorrectly reply. We are interested in the maximum number of Byzantine servers in $B(t,t+T_r-\delta)$, so that the remaining ones belong to $B(t+T_r-\delta,t+T_r)$, which means that servers in $B(t+T_r-\delta,t+T_r)$ are in $Co(t,t+\delta)$ (considering the scenario $S^*$). Thus, considering  that in the $(\Delta,CUM)$ model we consider $t_BB(s_b)=t+T_r-\epsilon$ ($\epsilon>0$ and arbitrarily small) then we consider the maximum number of \vir{jumps} there could be in the time interval $[t+\delta,t+T_r-\epsilon]$. Thus $min\tilde{CB}(t,t+T_r)=\ceil{\frac{t+T_r-\epsilon-t-\delta}{\Delta}}=\ceil{\frac{T_r-\epsilon-\delta}{\Delta}}$. If $MaxCu(t)=0$ or $\gamma>\Delta$ then it has no sense to consider the $(\Delta S, CUM)$ worst case scenario that aims to maximize cured servers. Thus in this case we consider the $(\Delta S,CAM)$ worst case scenario,   $min\tilde{CB}=\mathcal{R}(Max\tilde{B}(t,t+T_r)-Max\tilde{B}(t,t+T_r-\delta))$, concluding the proof.
\renewcommand{\toto}{l:minCBCUMGenerale}
\end{proofL}

\begin{lemma}\label{l:minCBCCUMGenerale}
Let us consider a time interval $[t,t+T_r], T_r\geq 2\delta$ then in the $(\Delta S,CUM)$ model then if $maxCu(t)>0$ $min\tilde{CBC}=\ceil{\frac{T_r-\epsilon-\delta}{\Delta}}+\mathcal{R}(\ceil{\frac{T_r}{\Delta}}-\ceil{\frac{\gamma-\delta}{\Delta}})+(MaxCu(t)-MaxSil(t,t+T_r))$, otherwise $min\tilde{CBC}$ assumes the same values as in the $(\Delta S, CAM)$ case. 
\end{lemma}

\begin{proofL}
By definition $min\tilde{CBC}(t,t+T_r)=min\tilde{CB}(t,t+T_r)+min\tilde{BC}(t,t+T_r)$. 
From Lemma \ref{l:minCBCUMGenerale}, if $maxCu(t)>0$ or $\Delta>\gamma$ then in the $(\Delta S,CUM)$ model $min\tilde{CB}=\ceil{\frac{T_r-\epsilon-\delta}{\Delta}}$ otherwise $min\tilde{CB}=\mathcal{R}(Max\tilde{B}(t,t+T_r)-Max\tilde{B}(t,t+T_r-\delta))$.\\
$min\tilde{BC}(t,t+T_r)$ is the minimum number of servers that incorrectly reply and then, before $t+T_r-\delta$ become correct so that are able to correctly reply in time such that their reply is delivered. In the $(\Delta S,CUM)$ model servers may incorrectly reply because affect by a mobile agent or because in a cured state. In the first case, a server is able to correctly reply if it become correct before $t+T_r-\delta$ (the reply message needs at most $\delta$ time to be delivered). Thus we consider the maximum number of servers that can be affected in the period $t+T_r-\gamma-\delta,t+T_r$, which is $\ceil{\frac{\gamma+\delta}{\Delta}}$.
Thus, among the Byzantine servers (i.e., $Max\tilde{B}(t,t+T)$) we consider servers not affected in the time interval $[t+T_r-\gamma+\delta, t+T_r]$. In other words such servers have $\gamma$ time to became correct and $\delta$ time to reply before the end of the operation. Thus $Max\tilde{B}(t,t+T_r)-Max(t+T_r-\gamma+\delta, t+T_r)$. Again we can not have a negative number, so it becomes $\mathcal{R}(\ceil{\frac{T_r}{\Delta}-\frac{\gamma-\delta}{\Delta}})$.
Concerning servers that incorrectly reply when in a cured state, we are interested in servers that correctly reply after in time such that the reply is delivered by the client, i.e., they are not silent. This number is easily computable, $MaxCu(t)-MaxSil(t,t+T_r)$. Thus $min\tilde{BC}(t,t+2\delta)=(MaxCu(t)-MaxSil(t,t+T_r))$.
Summing up if $maxCu(t)>0$ or $\Delta>\gamma$, then $min\tilde{CBC}=\ceil{\frac{T_r-\epsilon-\delta}{\Delta}}+\mathcal{R}(\ceil{\frac{T_r}{\Delta}}-\ceil{\frac{\gamma-\delta}{\Delta}})+(MaxCu(t)-MaxSil(t,t+2\delta))$, otherwise $min\tilde{CBC}$ assumes the same values as in the $(\Delta S, CAM)$ model, which concludes the proof.
\renewcommand{\toto}{l:minCBCCUMGenerale}
\end{proofL}

\begin{table*}[]
\centering
\caption{Values for a general \op{read} operation that terminates after $T_r$ time.}
\label{t:tabelloneGenerale}
\begin{tabular}{|c|c|c|c|}
\hline
 & $Max\tilde{B}(t,t+T_r)$ & $MaxCu(t)$ & $MaxSil(t,t+T_r)$  \\
 \hline
$(\Delta S,CAM)$ & $\ceil{\frac{T_r}{\Delta}}+1$ & $\mathcal{R}(\ceil{\frac{\gamma-\Delta+\epsilon}{\Delta}})$ & $\mathcal{R}(\ceil{\frac{\gamma-\Delta +\epsilon-T_r +\delta}{\Delta}})$  \\
\hline
$(\Delta S,CUM)$ & $\ceil{\frac{T_r}{\Delta}}+1$ & $ \mathcal{R}( \ceil{\frac{T_r - \epsilon - \ceil{\frac{T_r}{\Delta}}\Delta +\gamma}{\Delta}})$ & $\ceil{\frac{\gamma+\delta - \epsilon - \ceil{\frac{T_r}{\Delta}}\Delta }{\Delta}}$  \\
\hline
\end{tabular}
\begin{tabular}{|c|c|}
\hline
 & $min\tilde{CBC}(t,t+T_r)$ \\
 \hline
$(\Delta S,CAM)$  & $\mathcal{R}(\ceil{\frac{T_r}{\Delta}}-\ceil{\frac{\delta}{\Delta}})+\mathcal{R}(\ceil{\frac{T_r-\gamma-\delta}{\Delta}})$ \\
\hline
$(\Delta S,CUM)$ &  $\ceil{\frac{T_r-\epsilon-\delta}{\Delta}}$\footnote{if $maxCu(t)>0$ otherwise is the same value of $min\tilde{CBC}(t,t+T_r)$ in the $(*,CAM)$ model}$+\mathcal{R}(\ceil{\frac{T_r}{\Delta}}-\ceil{\frac{\gamma+\delta}{\Delta}})+(MaxCu(t)-MaxSil(t,t+T_r))$ \\
\hline
\end{tabular}
\end{table*}

In Table \ref{t:tabelloneGenerale} are reported all the results found so far for $(\Delta S, *)$ models.

Such results have been proved considering $f=1$. Extending such results to scenario for $f>1$ is straightforward in the $(\Delta S, *)$ model. The extension to $f>1$ in the $(ITB, *)$ and $(ITU, *)$ models is less direct. What is left to prove is that the results found for $f=1$ can be applied to all other models in which mobile agents move independently from each other. In the following Lemma we employ $*$ to indicate that the result holds for $*$ assuming consistently the value $CAM$ or $CUM$.

\begin{lemma}\label{l:mappingAsynchr}
Let $n_{*_{LB}}\leq \alpha_{*} (\Delta,\delta,\gamma)f$ be the impossibility results holding in the $(\Delta S,*)$ model for $f=1$. 
If there exists a tight protocol $\mathcal{P}_{reg}$ solving the safe register for $n \geq \alpha_{*}  (\Delta,\delta,\gamma)f+1$ ($f\geq 1$) then all the Safe Register impossibility results that hold in the $(\Delta S, *)$ models hold also in the $(ITB,*)$ and $(ITU,*)$ models.
\end{lemma}

\begin{proofL}
Let us consider the scenario $S^*$ for $f=1$ and a \op{read} operation time interval $[t,t+T_r]$, $t\geq 0$. Depending on the value of $t$ there can be different (but finite) read scenarios, $rs_1,rs_2, \dots, rs_s$.
By hypothesis there exists $\mathcal{P}_{reg}$ solving the safe register for  $n \geq \alpha_* f (\Delta,\delta,\gamma)+1$ then among the read scenarios $\mathcal{RS}=\{rs_1,rs_2, \dots, rs_s\}$ all the possible worst case scenarios $\{wrs_1,\dots, wrs_w\} \subseteq \mathcal{RS}$ hold for $n=\alpha_* (\Delta,\delta,\gamma)f$ (meaning that $\mathcal{P}_{reg}$ does not exist). We can say that those worst scenarios are equivalent in terms of replicas, i.e., for each $wsr_k$ is it possible to build an impossibility run if $n = \alpha_* (\Delta,\delta,\gamma)$ but $\mathcal{P}_{reg}$ works if $n = \alpha_* (\Delta,\delta,\gamma)+1$ (if we consider $f=1$).
Let us now consider $(\Delta S, *)$ for $f>1$. In this case, mobile agents move all together, thus the same $wrs_k$ scenario is reproduced $f$ times. For each $wrs_k$ scenario is it possible to build an impossibility run if $n = \alpha_{*}(\Delta,\delta,\gamma)f$, i.e., $\alpha_{*}(\Delta,\delta,\gamma)-1$ non Byzantine servers are not enough to cope with $1$ Byzantine server, then it is straightforward that $\alpha_*(\Delta,\delta,\gamma)-f$ non Byzantine servers are not enough to cope with $f$ Byzantine servers, the same scenario is reproduced $f$ times.\\
In the case of unsynchronized movements (ITB and ITU)
we consider $\Delta=min $ $\{\Delta_1, \dots, \Delta_f\}$. Each mobile agent generates a different read scenarios, those scenario can be up to $f$. As we just stated, if $\mathcal{P}_{reg}$ exists, those worst case scenarios are equivalent each others in terms of replicas. Since all the worst case scenarios are equivalent in terms of replicas, thus impossibility results holding for mobile agents moving together hold also for mobile agent moving in an uncoordinated way.  
\renewcommand{\toto}{l:mappingAsynchr}
\end{proofL}

In \cite{BDPT16c},  for $n \geq \alpha_{*}  (\Delta,\delta,\gamma)f+1$ ($f\geq 1$), it has been presented a tight protocol $\mathcal{P}_{reg}$ that solves the Regular Register problem whose bounds match the safe register lower bounds. Thus the next corollary follows.

\begin{corollary}\label{c:mappingEffettivo}
Let $n_{*_{LB}}\leq \alpha_{*} (\Delta,\delta,\gamma)f$ be the impossibility results holding in the $(\Delta S,*)$ model for $f=1$. All the Safe Register impossibility results hold also in the $(ITB,*)$ and $(ITU,*)$ models.
\end{corollary}

\section{Upper Bounds}\label{sec:algorithm}
In this section, we present an overview of the optimal protocols that implement a SWMR Regular Register in a round-free synchronous system respectively for $(ITB, CAM)$ and $(ITB, CUM)$ instances of the proposed MBF model. 

Following the same approach we used in \cite{BDPT16c} for the $(\Delta S, CAM)$ model, our solution is based on the following two key points: (1) we implement a ${\sf maintenance}()$ operation, in this case executed on demand; 
(2) we implement ${\sf read}()$ and ${\sf write}()$ operations following the classical quorum-based approach. The size of the quorum needed to carry on the operations, and consequently the total number of servers required by the computation, is dependent on the time to terminate the ${\sf maintenance}()$ operation, $\delta$ and $\Delta$ (see Table \ref{t:tableC*m}).  
The difference with respect $(\Delta S, CAM)$ model is that the time at which mobile agents move is unknown. Notice that each mobile $ma_i$ agent has it own $\Delta_i$. Since we do not have any other information we consider $\Delta = min \{\Delta_1, \dots, \Delta_f\}$.

\begin{table}[]
\centering
\caption{Parameters for $\mathcal{P}_{Reg}$ Protocol in the $(ITB, CAM)$ and $(ITB, CUM)$ models, minimum number of replicas, and minimum expected occurrence of correct values. 
}
\label{t:tableC*m}
\begin{tabular}{l|l|ll}
                             \textbf{(CAM, ITB)}                           & \multicolumn{2}{l}{} &  \\ \hline
\multirow{2}{*}{$2\delta \leq \Delta \textless 3 \delta$} & \multicolumn{2}{l}{$n_{CAM}$}          & 4f+1                   \\
                                                        & \multicolumn{2}{l}{$\#reply_{CAM}$}      & 2f+1                      \\
                                                        \hline
\multirow{2}{*}{$\delta \leq \Delta \textless 2 \delta$} & \multicolumn{2}{l}{$n_{CAM}$}          & 6f+1                      \\
                                                        & \multicolumn{2}{l}{$\#reply_{CAM}$}      & 3f+1                  
\end{tabular}\hspace{.5cm}
\begin{tabular}{l|l|ll}
                             \textbf{(CUM, ITB)}                    & \multicolumn{2}{l}{} &  \\ \hline
\multirow{2}{*}{$2\delta \leq \Delta \textless 3 \delta$} & \multicolumn{2}{l}{$n_{CUM}$}          & 7f+1                     \\
                                                        & \multicolumn{2}{l}{$\#reply_{CUM}$}      & 4f+1                 \\
                                                        \hline
\multirow{2}{*}{$\delta \leq \Delta \textless 2 \delta$} & \multicolumn{2}{l}{$n_{CUM}$}          & 12f+1                    \\
                                                        & \multicolumn{2}{l}{$\#reply_{CUM}$}      & 7f+1                 
\end{tabular}

\end{table} 


\noindent{\bf The ${\sf maintenance}()$ operation for $(ITB, CAM)$ model.} This operation is executed by servers on demand (request-reply) when the oracle notifies them that are in a cured state. Notice that in the $(*,CAM)$ models servers know when a mobile agent leaves them, thus depending on such knowledge they execute different actions. In particular, if a server $s_i$ is not in a cured state then it does nothing, it just replies to \msg{echo\_req} messages. 
Otherwise, if a server $s_i$ is in a cured state it first cleans its local variables and ${\sc broadcast}$ to other servers a request. Then, after $2\delta$ time units it removes values that may come from servers that were Byzantine before the \op{maintenance} and updates its state by checking the number of occurrences of each value received from the other servers. Contrarily to the $(\Delta S, CAM)$ case, a cured server notifies to all servers that it was Byzantine in the previous $\delta$ time period. This is done invoking the ${\sf awareAll}$ function that broadcasts a default value $\bot$ after $\delta$ time a server discovered to be in a cured state. This is done to prevent a cured server to collect \vir{slow} replies coming from servers that were affected before the execution of the \op{maintenance} operation. In this model, the curing time $\gamma\leq 2\delta$.\\

\noindent{\bf The ${\sf maintenance}()$ operation for $(ITB, CUM)$ model.} In this case servers are not aware of their failure state, thus they have to run such operation even if they are correct or cured. In addition, in the $(ITB, CUM)$ model, the moment at which mobile agents move is not known, thus as for the $(ITB,CAM)$ case, a request-reply pattern is used to implement the \op{maintenance} operation. Such operation is executed by servers every $2\delta$ times. In this case, to prevent a cured server to collect \vir{slow} replies coming from servers that were affected before the execution of the \op{maintenance} operation, a server choses a random number to associate to such particular \op{maintenance} operation instance \footnote{Is it out of the scope of this work to describe such function, we assume that Byzantine server can not predict the random number chosen next.}, broadcast the \msg{echo\_req} message and waits $2\delta$ before restart ing the operation. When there is a value whose occurrence overcomes the $\#echo_{CUM}$ threshold, such value is stored at the server side.  \\
Notice that, contrarily to all the previous models, servers are not aware of their failure state and do not synchronize the \op{maintenance} operation with each other. The first consequence is  that a mobile agent may leave a cured server running such operation with garbage in server variables, making the operation unfruitful. Such server has to wait $2\delta$ to run again the \op{maintenance} operation with clean variables, so that next time it will be effective, which implies $\gamma \leq 4\delta$.  \\

\noindent {\bf The ${\sf write}$ operation.} 
To write a new value $v$, the writer increments its sequence number $csn$ and propagates $v$ and $csn$ to all servers via a {\sc write} messages. Then, it waits for $\delta$ time units (the maximum message transfer delay) before returning. 
When a server $s_i$ delivers a {\sc write}, it updates its local variables and sends a {\sc reply} message to all clients that are currently reading to allow them to complete their ${\sf read}$ operation.\\

\noindent{\bf The ${\sf read}$ operation.} When a client wants to read, it broadcasts a {\sc read} request to all servers and then waits $2\delta$ time (\emph{i.e.}, one round trip delay) to collect replies.
When it is unblocked from the wait statement, it selects a value $v$ occurring enough number of times (see $\#reply_{C*M}$ from Table~\ref{t:tableC*m}) from the replies set, sends an acknowledgement message to servers to inform that its operation is now terminated and returns $v$ as result of the operation. When a server $s_i$ delivers a {\sc read}$(j)$ message from client $c_j$, it first puts its identifier in the pending read set to remember that $c_j$ is reading and needs to receive possible concurrent updates and it sends a reply back to $c_j$. 

\begin{table*}[t]
\centering
	\scriptsize
	\begin{tabular}{| c | c | c | c | }
		\hline
		$ {k=\ceil{\frac{2\delta}{\Delta}}\geq 1}$& $ {n_{CUM} \ge (5k+2)f+1}$ &  $ {\# reply_{CUM} \geq (3k+1)f+1}$ & $\# echo_{CUM}\geq (3k)+1f$  \\
		\hline
		$k=2$  & $12f+1$ &  $7f+1$ &  $6f+1$  \\
		\hline
		$k=1$  & $7f+1$ &  $4f+1$ &  $4f+1$   \\
		\hline
	\end{tabular}
	\caption{Parameters for $\mathcal{P}_{Rreg}$ Protocol for the $(ITB,CUM)$ model.}\label{tab:summaryITBCUM}
\end{table*}

\medskip

\subsection{$\mathcal{P}_{reg}$ in the $(ITB, CAM) model$}
The protocol $\mathcal{P}_{reg}$ for the $(ITB, CAM)$ model is described in Figures \ref{fig:stateMaintenanceProtocolITBCAM} - \ref{fig:readProtocolITBCAM}, which present the ${\sf maintenance}()$, ${\sf write}()$, and ${\sf read}()$ operations, respectively. \\

\noindent{\bf Local variables at client $c_i$.} Each client $c_i$ maintains a set $reply_i$ that is used during the ${\sf read}()$ operation to collect the three tuples $\langle j, \langle v, sn \rangle \rangle$ sent back from servers. In particular $v$ is the value, $sn$ is the associated sequence number and $j$ is the identifier of server $s_j$ that sent the reply back. Additionally, $c_i$ also maintains a local sequence number $csn$ that is incremented each time it invokes a ${\sf write}()$ operation and is used to timestamp such operations monotonically.\\


\noindent{\bf Local variables at server $s_i$.} Each server $s_i$ maintains the following local variables (we assume these variables are initialized to zero, false or empty sets according their type):

\begin{itemize}

	\item { $V_i$: an ordered set containing $d$ tuples $\langle v, sn \rangle$, where $v$ is a value and $sn$ the corresponding sequence number. Such tuples are ordered incrementally according to their $sn$ values. The function ${\sf insert}(V_i, \langle v_k, sn_k \rangle)$ places the new value in $V_i$ according to the incremental order and, if there are more than $d$ values, it discards from $V_i$ the value associated to the lowest $sn$.}
	
	\item $pending\_read_i$: set variable used to collect identifiers of the clients that are currently reading.
	
	\item $cured_i$:  boolean flag updated by the ${\sf cured\_state}$ oracle. In particular, such variable is set to ${\sf true}$ when $s_i$ becomes aware of its cured state and it is reset during the algorithm when $s_i$ becomes correct. 
	
	\item $echo\_vals_i$ and $echo\_read_i$: two sets used to collect information propagated through {\sc echo} messages. The first one stores tuple $\langle j, \langle v, sn \rangle \rangle$ propagated by servers just after the mobile Byzantine agents moved, while the second stores the set of concurrently reading clients in order to notify cured servers and expedite termination of ${\sf read}()$.
	
	\item $curing_i$: set used to collect servers running the \op{maintenance} operation. Notice, to keep the code simple we do not explicitly manage how to empty such set since has not impact on safety properties.
	
\end{itemize}

In order to simplify the code of the algorithm, let us define the following functions:

\begin{itemize}
\item ${\sf select\_d\_pairs\_max\_sn} (echo\_vals_i)$: this function ta\-kes as input the set $echo\_vals_i$ and returns, if they exist, three tuples $\langle v, sn \rangle$, such that there exist at least $\#echo_{CAM}$ occurrences in $echo\_vals_i$ of such tuple. If more than three of such tuple exist, the function returns the tuples with the highest sequence numbers. 

\item ${\sf select\_value}(reply_i)$: this function takes as input the $reply_i$ set of replies collected by client $c_i$ and returns the pair $\langle v, sn \rangle$ occurring at least $\# reply_{CAM}$ times (see Table \ref{tab:summaryITBCAM}). If there are more pairs satisfying such condition, it returns the one with the highest sequence number.

\item ${\sf delete\_cured\_values(echo\_vals)}$: this function takes as input $echo\_vals_i$ and removes from $fw\_vals_i$ all values coming from servers that sent an \msg{echo} message containing $\bot$. 

\end{itemize}


\begin{figure*}[t]
\centering
\fbox{
\begin{minipage}{0.4\textwidth}
\scriptsize
\resetline
\begin{tabbing}
aaaA\=aA\=aA\=aaaA\kill

{\bf function} ${\sf awareAll}()$:\\

\line{M-ITBCAM-01} \>${\sf broadcast}$ {\sc echo}$(i, \bot)$\\

\line{M-ITBCAM-02} \> {\bf wait}($\delta$);\\

\line{M-ITBCAM-03} \> ${\sf broadcast}$ {\sc echo}$(i, \bot)$\\

~------------------------------------------------------------------------------------------------------\\

{\bf operation} ${\sf maintenance}()$ {\bf executed while} {\sc(true)}  :\\

\line{M-ITBCAM-04} \> $cured_i \leftarrow {\sf report\_cured\_state}()$; \\

\line{M-ITBCAM-05} \> {\bf if} \= $(cured_i)$ {\bf then}\\

\line{M-ITBCAM-06}\>\> $cured_i \leftarrow {\sc false}$;\\

\line{M-ITBCAM-07}\>\> $curing\_state_i \leftarrow {\sc true}$;\\

\line{M-ITBCAM-08}\>\> $V_i \leftarrow \emptyset$; $echo\_vals_i \leftarrow \emptyset$; $pending\_read_i \leftarrow \emptyset$;$curing_i \leftarrow \emptyset$;\\

\line{M-ITBCAM-10}\>\> ${\sf broadcast}$ {\sc echo\_req}$(i)$;\\

\line{M-ITBCAM-10b}\>\> ${\sf awareAll}()$;\\

\line{M-ITBCAM-11} \>\> {\bf wait}$(2\delta)$;\\

\line{M-ITBCAM-12} \>\> {\sf delete\_cured\_values(echo\_vals)};\\

\line{M-ITBCAM-13}\>\> ${\sf insert}(V_i, {\sf select\_three\_pairs\_max\_sn} (echo\_vals_i) )$;\\

\line{M-ITBCAM-14} \>\>{\bf for each} \= $(j \in (curing_i))$ {\bf do}\\

\line{M-ITBCAM-15} \>\>\> ${\sf send}$ {\sc echo} $(i, V_i )$ to $s_j$;\\

\line{M-ITBCAM-16} \>\>{\bf endFor}\\

\line{M-ITBCAM-17} \>\> $curing\_state_i \leftarrow {\sf false}$;\\


\line{} \> {\bf endIf}\\
%
%
%
%

~------------------------------------------------------------------------------------------------------\\

{\bf when} \=  {\sc echo} $(j, V_j)$ \= is  ${\sf received}$: \\ 

\line{M-ITBCAM-18} \>{\bf for each} \= $(\ang{v}{sn} \in V_j$ {\bf do}\\

\line{M-ITBCAM-19} \>\> $echo\_vals_i \leftarrow echo\_vals_i \cup \ang{v}{sn}_j$;\\

\line{M-ITBCAM-20} \>{\bf endFor}\\

~------------------------------------------------------------------------------------------------------\\

{\bf when} \=  {\sc echo\_req} $(j)$ \= is  ${\sf received}$: \\ 

\line{M-ITBCAM-21} \> $curing_i \leftarrow curing_i \cup {j}$;\\

\line{M-ITBCAM-22} \>{\bf if} \= $(V_i \neq \emptyset)$\\

\line{M-ITBCAM-23}\>\> ${\sf send}$ {\sc echo}$(i, V_i)$;\\

\line{M-ITBCAM-24}\> {\bf endif}

\end{tabbing}
\normalsize
\end{minipage}
}
\caption{$\mathcal{A}_{M}$ algorithm implementing the ${\sf maintenance}()$ operation (code for server $s_i$) in the $(ITB, CAM)$ model.}
\label{fig:stateMaintenanceProtocolITBCAM}   
\end{figure*}

\noindent{\bf The ${\sf maintenance}()$ operation.} Such operation is executed by servers on demand when the oracle notifies them that are in a cured state. Notice that in the $(*,CAM)$ models servers knows when a mobile agent leaves them, thus depending on such knowledge they execute different actions. In particular, if a server $s_i$ is not in a cured state then it does nothing, it just replies to \msg{echo\_req} messages. 
Otherwise, if a server $s_i$ is in a cured state it  first cleans its local variables and ${\sc broadcast}$ to other servers an echo request then, after $2\delta$ time units it removes value that may come from servers that were Byzantine before the \op{maintenance} and updates its state by checking the number of occurrences of each pair $\langle v, sn \rangle$ received with {\sc echo} messages. In particular, it updates $V_i$ invoking the ${\sf select\_three\_pairs\_max\_sn} (echo\_vals_i)$ function that populates $V_i$ with $d$ tuples $\langle v, sn \rangle$. 
At the end it assigns ${\sc false}$ to $cured_i$ variable, meaning that it is now correct and the $echo\_vals_i$ can now be emptied.
Contrarily to the $(\Delta S, CAM)$ case, cured server notifies to all that it has been Byzantine in the previous $\delta$ time period. This is done invoking the ${\sf awareAll}$ function that broadcast a default value $\bot$ after $\delta$ time that a server discovered to be in a cured state. 


\begin{figure*}[t]
\centering
\fbox{
\begin{minipage}{0.4\textwidth}
\scriptsize
\resetline

\begin{tabbing}
aaaA\=aA\=aA\=aaaA\kill
========= {\bf Client code} ==========\\

{\bf operation} ${\sf write}(v)$: ~~~~~~~~~~~~~~~~~~~~~~~~~~~~~~~~~ \\


\line{W-ITBCCAM-01} \> $csn \leftarrow csn+1$;\\

\line{W-ITBCCAM-02} \> ${\sf broadcast}$ {\sc write}$(v, csn)$;\\

\line{W-ITBCCAM-03} \> {\bf wait} $(\delta)$;\\

\line{W-ITBCCAM-04} \> {\bf return} ${\sf write\_confirmation}$;

\end{tabbing}

\begin{tabbing}
========= {\bf Server code} ==========\\
aaaA\=aA\=aA\=aaA\kill

{\bf when} \=  {\sc write}$(v, csn)$ \= is  ${\sf received}$: \\ 

\line{W-ITBSCAM-01} \> ${\sf insert}(V_i, \langle v, csn \rangle)$;\\

\line{W-ITBSCAM-02} \>  {\bf for} \= {\bf each} $j \in (pending\_read_i)$ {\bf do}\\

\line{W-ITBSCAM-03} \>  \> ${\sf send}$ {\sc reply} $(i, \{\langle v, csn \rangle\} )$;\\

\line{W-ITBSCAM-04} \> {\bf endFor}\\

\line{W-ITBSCAM-05} \>  {\bf for} \= {\bf each} $j \in (curing_i )$ {\bf do}\\

\line{W-ITBSCAM-06} \>  \> ${\sf send}$ {\sc echo} $(i, V_i)$;\\

\line{W-ITBSCAM-07} \> {\bf endFor}\\

\end{tabbing}
\end{minipage}
}
\caption{$\mathcal{A}_{W}$ algorithm implementing the ${\sf write}(v)$ operation in the $(ITB, CAM)$ model.}
\label{fig:writeProtocolITBCAM} 
\end{figure*}

%

%
%
%
%
\noindent {\bf The ${\sf write}()$ operation.} 
When the writer wants to write a value $v$, it increments its sequence number $csn$ and propagates  $v$ and $csn$ to all servers. Then it waits for $\delta$ time units (the maximum message transfer delay) before returning. 

When a server $s_i$ delivers a {\sc write}, it updates its local variables and sends a {\sc reply}$()$ message to all clients that are currently reading (clients in $pending\_read_i$) to notify them about the concurrent \op{write} operation and to each server executing the \op{maintenance} operation (servers in $curing_i$).

\noindent{\bf The ${\sf read}()$ operation.} When a client wants to read, it broadcasts a {\sc read}$()$ request to all servers and  waits $2\delta$ time (i.e., one round trip delay) to collect replies.
When it is unblocked from the wait statement, it selects a value $v$ invoking the ${\sf select\_value}$ function on $reply_i$ set, sends an acknowledgement message to servers to inform that its operation is now terminated and returns $v$ as result of the operation.

When a server $s_i$ delivers a {\sc read}$(j)$ message from client $c_j$ it first puts its identifier in the set $pending\_read_i$ to remember that $c_j$ is reading and needs to receive possible concurrent updates, then $s_i$ checks if it is in a cured state and if not, it sends a reply back to $c_j$. Note that, the {\sc reply}$()$ message carries the set $V_i$.

When a {\sc read\_ack}$(j)$ message is delivered, $c_j$ identifier is removed from both $pending\_read_i$ set as it does not need anymore to receive updates for the current ${\sf read}()$ operation.

\begin{figure}[h!]
\centering

\fbox{
\begin{minipage}{0.4\textwidth}
\scriptsize
\resetline
\begin{tabbing}
========= {\bf Client code} ==========\\
aaaA\=aA\=aA\=aaaA\kill

{\bf operation} ${\sf read}()$: ~~ \\


\line{R-ITBCCAM-01} \> $reply_i \leftarrow \emptyset$;\\

\line{R-ITBCCAM-02} \> ${\sf broadcast}$ {\sc read}$(i)$;\\

\line{R-ITBCCAM-03} \> {\bf wait }$(2\delta)$;\\

\line{R-ITBCCAM-04} \> $\langle v, sn \rangle \leftarrow {\sf select\_value}(reply_i)$;\\

\line{R-ITBCCAM-05} \> ${\sf broadcast}$ {\sc read\_ack}$(i)$;\\

\line{R-ITBCCAM-06} \> {\bf return} $v$;\\

~-----------------------------------------------------------------------\\

{\bf when} \=  {\sc reply} $(j, V_j)$ \= is  ${\sf received}$: \\ 

\line{R-ITBCCAM-07} \> {\bf for each}  \= $(\langle v, sn \rangle \in V_j)$ {\bf do}\\

\line{R-ITBCCAM-08} \>\> $reply_i \leftarrow reply_i \cup \{ \langle j, \langle v, sn \rangle \rangle \}$;\\

\line{R-ITBCCAM-09} \> {\bf endFor}\\

\end{tabbing}

\begin{tabbing}
========= {\bf Server code} ==========\\
aaaA\=aA\=aA\=aaA\kill

{\bf when} \=  {\sc read} $(j)$ \= is  ${\sf received}$: \\ 

\line{R-ITBSCAM-01} \> $pending\_read_i \leftarrow pending\_read_i \cup \{j\}$;\\

\line{R-ITBSCAM-02} \> {\bf if} \= $(V_i \neq \emptyset)$\\

\line{R-ITBSCAM-03} \>\> {\bf then} \= ${\sf send}$ {\sc reply} $(i, V_i)$;\\

\line{R-ITBSCAM-04} \> {\bf endif}\\


~-----------------------------------------------------------------------\\


{\bf when} \=  {\sc read\_ack} $(j)$ \= is  ${\sf received}$: \\ 

\line{R-ITBSCAM-05}\>$pending\_read_i \leftarrow pending\_read_i \setminus \{j\}$;\\


\end{tabbing}

\normalsize
\end{minipage}

}
\caption{$\mathcal{A}_{R}$ algorithm implementing the ${\sf read}()$ operation in the $(ITB, CAM)$ model.}
\label{fig:readProtocolITBCAM}   
\end{figure}

\subsection{$\mathcal{P}_{reg}$ in the $(ITB, CUM) model$}
\noindent\textbf{$\mathcal{P}_{reg}$ Detailed Description}
The protocol $\mathcal{P}_{reg}$ for the $(ITB, CUM)$ model is described in Figures \ref{fig:stateMaintenanceProtocolITBCUM} - \ref{fig:readProtocolITBCUM}, which present the ${\sf maintenance}()$, ${\sf write}()$, and ${\sf read}()$ operations, respectively. Table \ref{tab:summaryITBCUM} reports the parameters for the protocol. In particular $n_{CUM}$ is the bound on the number of servers, $\#reply_{CUM}$ is minimum number of occurrences from different servers of a value to be accepted as a reply during a \op{read} operation and $\#echo_{CUM}$ is the minimum number of occurrences from different servers of a value to be accepted during the \op{maintenance} operation.\\

\noindent{\bf Local variables at client $c_i$.} Each client $c_i$ maintains a set $reply_i$ that is used during the ${\sf read}()$ operation to collect the three tuples $\langle j, \langle v, sn \rangle \rangle$ sent back from servers. In particular $v$ is the value, $sn$ is the associated sequence number and $j$ is the identifier of server $s_j$ that sent the reply back. Additionally, $c_i$ also maintains a local sequence number $csn$ that is incremented each time it invokes a ${\sf write}()$ operation and is used to timestamp such operations monotonically.\\


\noindent{\bf Local variables at server $s_i$.} Each server $s_i$ maintains the following local variables (we assume these variables are initialized to zero, false or empty sets according their type):

\begin{itemize}

	\item $V_i$: an ordered set containing $3$ tuples $\langle v, sn \rangle$, where $v$ is a value and $sn$ the corresponding sequence number. Such tuples are ordered incrementally according to their $sn$ values.
	
	\item $V_{safe_j}$: this set has the same characteristic as $V_j$. The  ${\sf insert}(V_{safe_i}, \langle v_k, sn_k \rangle)$ function places the new value in $V_{safe_i}$ according to the incremental order and if dimensions exceed $3$ then it discards from $V_{safe_i}$ the value associated to the lowest $sn$.
	
	\item $W_i$: is the set where servers store values coming directly from the writer, associating to it a timer, $\langle v, sn, timer \rangle$. Values from this set are deleted when the timer expires or has a value non compliant with the protocol.  
	
	\item $pending\_read_i$: set variable used to collect identifiers of the clients that are currently reading.
		
	\item $echo\_vals_i$ and $echo\_read_i$: two sets used to collect information propagated through {\sc echo} messages. The first one stores tuple $\langle j, \langle v, sn \rangle \rangle$ propagated by servers just after the mobile Byzantine agents moved, while the second stores the set of concurrently reading clients in order to notify cured servers and expedite termination of ${\sf read}()$.
	
	\item $curing_i$: set used to collect servers running the \op{maintenance} operation. Notice, to keep the code simple we do not explicitly manage how to empty such set since has not impact on safety properties.
	
\end{itemize}

In order to simplify the code of the algorithm, let us define the following functions:

\begin{itemize}
\item ${\sf select\_three\_pairs\_max\_sn} (echo\_vals_i)$: this function takes as input the set $echo\_vals_i$ and returns, if they exist, three tuples $\langle v, sn \rangle$, such that there exist at least $\# echo_{CUM}$ occurrences in $echo\_vals_i$ of such tuple. If more than three of such tuples exist, the function returns the tuples with the highest sequence numbers. 

\item ${\sf select\_value}(reply_i)$: this function takes as input the $reply_i$ set of replies collected by client $c_i$ and returns the pair $\langle v, sn \rangle$ occurring  occurring at least $\#reply_{CUM}$ times. If there are more pairs with the same occurrence, it returns the one with the highest sequence number.

\item ${\sf conCut}(V_i, V_{safe_i},W_i)$: this function takes as input three $3$ dimension ordered sets and returns another $3$ dimension ordered set. The returned set is composed by the concatenation of $V_{safe_i} \circ V_i \circ W_i$, without duplicates, truncated after the first $3$ newest values (with respect to the timestamp). e.g., $V_i=\{\ang{v_a}{1},\ang{v_b}{2}, \ang{v_c}{3}, \ang{v_d}{4}\}$ and $V_{safe_i}=\{\ang{v_b}{2}, \ang{v_d}{4}, \ang{v_f}{5}\}$ and $W_i=\emptyset$, then the returned set is $\{\ang{v_c}{3}, \ang{v_d}{4}, \ang{v_f}{5}\}$.

\end{itemize}


\begin{figure*}[t]
\centering
\fbox{
\begin{minipage}{0.4\textwidth}
\scriptsize
\resetline
\begin{tabbing}
aaaA\=aA\=aA\=aaaA\kill

{\bf operation} ${\sf timerCheck}(W_i)$ {\bf executed while} {\sc(true)}  :\\

\line{T-01}\> {\bf for each} \= $(\ang{\ang{v}{csn}}{timer}_j \in W_i $) {\bf do}\\

\line{T-02} \>\> {\bf if} \= $(Expires(timer) \wedge (timer >4\delta))$\\

\line{T-03} \>\>\> $W_i \leftarrow W_i \setminus \ang{\ang{v}{csn}}{timer}_j $;\\

\line{T-05} \>\> {\bf endif}\\

\line{T-06}\>{\bf endFor}\\
~-------------------------------------------------------------------------------------------------------------\\


{\bf operation} ${\sf maintenance}()$ {\bf executed while} {\sc(true)}  :\\

\line{M-ITBCUM-01} \> $echo\_vals_i\leftarrow \emptyset$; $V_i\leftarrow V_{safe_i}$; $V_{safe} \leftarrow \emptyset$; \\

\line{M-ITBCUM-02}\> $rand \leftarrow$ {\sf new\_rand}$()$;\\

\line{M-ITBCUM-03}\> ${\sf broadcast}$ {\sc echo\_req}$(i, rand)$;\\

\line{M-ITBCUM-04} \> {\bf wait}$(2\delta)$;\\

~------------------------------------------------------------------------------------------------------\\

{\bf when} ${\sf select\_three\_pairs\_max\_sn} (echo\_vals_i)\neq \bot$\\

\line{M-ITBCUM-06}\> ${\sf insert}(V_{safe_i}, {\sf select\_three\_pairs\_max\_sn} (echo\_vals_i) )$;\\

\line{M-ITBCUM-07} \>{\bf for each} \= $(j \in (pending\_read_i \cup echo\_read_i))$ {\bf do}\\

\line{M-ITBCUM-08} \>\> ${\sf send}$ {\sc reply} $(i, V_{safe} )$ to $c_j$;\\

\line{M-ITBCUM-09} \>{\bf endFor}\\

~-------------------------------------------------------------------------------------------------------------\\

{\bf when} \=  {\sc echo} $(j, S, pr,r)$ \= is  ${\sf received}$: \\ 

\line{M-ITBCUM-10} \> {\bf if} \= $(rand=r)${\bf then}:\\

\line{M-ITBCUM-11} \>\> $echo\_vals_i \leftarrow echo\_vals_i \cup \langle v, sn \rangle_j$;\\

\line{M-ITBCUM-12} \>\> $echo\_read_i \leftarrow echo\_read_i \cup pr$;\\

\line{M-ITBCUM-13}\> {\bf endIf}\\

~------------------------------------------------------------------------------------------------------\\

{\bf when} \=  {\sc echo\_req} $(j,r)$ \= is  ${\sf received}$: \\ 

\line{M-ITBCUM-14} \> $Set_i \leftarrow \emptyset$;\\

\line{M-ITBCUM-15}\> {\bf for each}$ \ang{\ang{v}{csn}}{epoch}_j \in W_i$ {\bf do};\\

\line{M-ITBCUM-16}\> $Set_i \leftarrow Set_i \cup \ang{v}{csn}_j$;\\

\line{M-ITBCUM-17}\>{\bf endFor} \\

\line{M-ITBCUM-18}\> ${\sf send}$ {\sc echo}$(i, V_i \cup Set_i,r)$ to $s_j$;\\

\end{tabbing}
\normalsize
\end{minipage}
}
\caption{$\mathcal{A}_{M}$ algorithm implementing the ${\sf maintenance}()$ operation (code for server $s_i$) in the $(ITB,CUM)$ model.}
\label{fig:stateMaintenanceProtocolITBCUM}   
\end{figure*}


%
\noindent{\bf The ${\sf maintenance}()$ operation.} Such operation is executed by servers every $2\delta$ times. Each time $s_i$ resets its variables, except for $W_i$ (that is continuously checked by the function {\sf timerCheck}$()$) and the content of $V_{safe_i}$, which overrides the content of $V_i$, before to be reset. Then $s_i$ choses a random number to associate to such particular \op{maintenance} operation instance \footnote{Is it out of the scope of this work to describe such function, we assume that Byzantine server can not predict the random number chosen next. The aim of such number is to prevent Byzantine servers to send reply to \op{maintenance} operations before their invocation, or, in other words, it prevents correct servers to accept those replies.}, broadcast the \msg{echo\_req} message and waits $2\delta$ before to restart the operation. In the meantime \msg{echo} messages are delivered and stored in the $echo\_vals_i$ set. When there is value $v$ whose occurrence overcomes the $\#echo_{CUM}$ threshold, such value is stored in $V_{safe_i}$ and a \msg{reply} message with $v$ is sent to current reader clients (if any).  \\
Notice that, contrarily to all the previous models, servers are not aware about their failure state and do not synchronize the \op{maintenance} operation with each other. The first consequence is a that a mobile agent may leave a cured server running such operation with garbage in server variables, making the operation unfruitful. Such server has to wait $2\delta$ to run again the \op{maintenance} operation with clean variables, so that next time it will be effective, which implies $\gamma \leq 4\delta$.


\begin{figure*}[t]
\centering
\fbox{
\begin{minipage}{0.4\textwidth}
\scriptsize
\resetline

\begin{tabbing}
aaaA\=aA\=aA\=aaaA\kill
========= {\bf Client code} ==========\\

{\bf operation} ${\sf write}(v)$: ~~~~~~~~~~~~~~~~~~~~~~~~~~~~~~~~~ \\


\line{W-CITBCUM-01} \> $csn \leftarrow csn+1$;\\

\line{W-CITBCUM-02} \> ${\sf broadcast}$ {\sc write}$(v, csn)$;\\

\line{W-CITBCUM-03} \> {\bf wait} $(\delta)$;\\

\line{W-CITBCUM-04} \> {\bf return} ${\sf write\_confirmation}$;

\end{tabbing}

\begin{tabbing}
========= {\bf Server code} ==========\\
aaaA\=aA\=aA\=aaA\kill

{\bf when} \=  {\sc write}$(v, csn)$ \= is  ${\sf received}$: \\ 

\line{W-SITBCUM-01} \> $W_i \leftarrow W_i \cup \ang{\ang{v}{csn}}{{\sf setTimer}(4\delta)\}}$;\\

\line{W-SITBCUM-02} \>  {\bf for} \= {\bf each} $j \in (pending\_read_i \cup echo\_read_i)$ {\bf do}\\

\line{W-SITBCUM-03} \> \>  ${\sf send}$ {\sc reply} $(i, \{\langle v, csn \rangle\} )$;\\

\line{W-SITBCUM-04} \> {\bf endFor}\\

\line{W-SITBCUM-05}\> ${\sf broadcast}$ {\sc echo}$(i, \ang{v}{csn})$;\\



\end{tabbing}
\end{minipage}
}
\caption{$\mathcal{A}_{W}$ algorithm implementing the ${\sf write}(v)$ operation in the $(ITB,CUM)$ model.}
\label{fig:writeProtocolITBCUM} 
\end{figure*}

%

%
%
%
%
\noindent {\bf The ${\sf write}()$ operation.} 
When the writer wants to write a value $v$, it increments its sequence number $csn$ and propagates  $v$ and $csn$ to all servers. Then it waits for $\delta$ time units (the maximum message transfer delay) before returning. 

When a server $s_i$ delivers a {\sc write} message, it updates $W_i$, associating to such value a timer $4\delta$. $4\delta$ it is a consequence of the double \op{maintenance} operation that a cured server has to run in order to be sure to be correct. Thus if a server is correct it keeps $v$ in $W_i$ during $4\delta$, which is enough for our purposes. On the other side a cured servers keeps a value (not necessarily coming from a \op{write} operation) no more than the time it is in a cured state, $4\delta$, which is safe. After storing $v$ in $W_i$, such value is inserted in {\sc reply}$()$ message to all clients that are currently reading (clients in $pending\_read_i$) to notify them about the concurrent \op{write} operation and to any server executing the \op{maintenance} operation (servers in $curing_i$). 

\noindent{\bf The ${\sf read}()$ operation.} When a client wants to read, it broadcasts a {\sc read}$()$ request to all servers and  waits $2\delta$ time (i.e., one round trip delay) to collect replies.
When it is unblocked from the wait statement, it selects a value $v$ invoking the ${\sf select\_value}$ function on $reply_i$ set, sends an acknowledgement message to servers to inform that its operation is now terminated and returns $v$ as result of the operation.

When a server $s_i$ delivers a {\sc read}$(j)$ message from client $c_j$ it first puts its identifier in the set $pending\_read_i$ to remember that $c_j$ is reading and needs to receive possible concurrent updates, then $s_i$  sends a reply back to $c_j$. Note that, in the {\sc reply}$()$ message is carried the result of {\sf conCut}$(V_i, V_{safe_i},W_i)$. In this case, if the server is correct then $V_i$ contains valid values, and $V_{safe_i}$ contains valid values by construction, since it comes from values sent during the current \op{maintenance}. If the server is cured, then $V_i$ and $W_i$ may contain any value. 
Finally, $s_i$ forwards a {\sc read\_fw} message to inform other servers about $c_j$ read request. This is useful in case some server missed the {\sc read}$(j)$ message as it was affected by mobile Byzantine agent when such message has been delivered.

When a {\sc read\_ack}$(j)$ message is delivered, $c_j$ identifier is removed from both $pending\_read_i$ set as it does not need anymore to receive updates for the current ${\sf read}()$ operation.


\begin{figure}[h!]
\centering

\fbox{
\begin{minipage}{0.4\textwidth}
\scriptsize
\resetline
\begin{tabbing}
========= {\bf Client code} ==========\\
aaaA\=aA\=aA\=aaaA\kill

{\bf operation} ${\sf read}()$: ~~ \\


\line{R-CCAM-01} \> $reply_i \leftarrow \emptyset$;\\

\line{R-CCAM-02} \> ${\sf broadcast}$ {\sc read}$(i)$;\\

\line{R-CCAM-03} \> {\bf wait }$(2\delta)$;\\

\line{R-CCAM-04} \> $\langle v, sn \rangle \leftarrow {\sf select\_value}(reply_i)$;\\

\line{R-CCAM-05} \> ${\sf broadcast}$ {\sc read\_ack}$(i)$;\\

\line{R-CCAM-06} \> {\bf return} $v$;\\

~-----------------------------------------------------------------------\\

{\bf when} \=  {\sc reply} $(j, V_j)$ \= is  ${\sf received}$: \\ 

\line{R-CCAM-07} \> {\bf for each}  \= $(\langle v, sn \rangle \in V_j)$ {\bf do}\\

\line{R-CCAM-08} \>\> $reply_i \leftarrow reply_i \cup \{ \langle j, \langle v, sn \rangle \rangle \}$;\\

\line{R-CCAM-09} \> {\bf endFor}\\

\end{tabbing} %
\begin{tabbing}
========= {\bf Server code} ==========\\
aaaA\=aA\=aA\=aaA\kill

{\bf when} \=  {\sc read} $(j)$ \= is  ${\sf received}$: \\ 

\line{R-SCUM-01} \> $pending\_read_i \leftarrow pending\_read_i \cup \{j\}$;\\

\line{R-SCUM-02} \> ${\sf send}$ {\sc reply} $(i, {\sf conCut}(V_i,V_{safe_i}, W_i))$;\\

\line{R-SCUM-03} \> ${\sf broadcast}$ {\sc read\_fw}$(j)$;\\

~-----------------------------------------------------------------------\\

{\bf when} \=  {\sc read\_fw} $(j)$ \= is  ${\sf received}$: \\ 

\line{R-SCUM-04} \> $pending\_read_i \leftarrow pending\_read_i \cup \{j\}$;\\

~-----------------------------------------------------------------------\\

{\bf when} \=  {\sc read\_ack} $(j)$ \= is  ${\sf received}$: \\ 

\line{R-SCUM-05}\>$pending\_read_i \leftarrow pending\_read_i \setminus \{j\}$;\\

\line{R-SCUM-06} \>$echo\_read_i \leftarrow echo\_read_i \setminus \{j\}$;

\end{tabbing}

\normalsize
\end{minipage}

}
\caption{$\mathcal{A}_{R}$ algorithm implementing the ${\sf read}()$ operation in the $(ITB,CUM)$ model.}
\label{fig:readProtocolITBCUM}   
\end{figure}


\section{Correctness}\label{sec:discusion}
\subsection{Correctness $(ITB, CAM)$}
To prove the correctness of $\mathcal{P}_{reg}$, we first show that the termination property is satisfied i.e, that ${\sf read}()$ and ${\sf write}()$ operations terminates.

\begin{lemma}\label{lem:wTermITBCAM}
	If a correct client $c_i$ invokes ${\sf write}(v)$ operation at time $t$ then this operation terminates at time $t+\delta$.
\end{lemma}

\begin{proofL}
	The claim follows by considering that a ${\sf write\_confirmation}$ event is returned to the writer client $c_i$ after $\delta$ time, independently of the behavior of the servers (see lines \ref{W-ITBCCAM-03}-\ref{W-ITBCCAM-04}, Figure \ref{fig:writeProtocolITBCAM}).
	\renewcommand{\toto}{lem:wTermITBCAM}
\end{proofL}

\begin{lemma}\label{lem:rTermITBCAM}
	If a correct client $c_i$ invokes ${\sf read}()$ operation at time $t$ then this operation terminates at time $t+2\delta$.
\end{lemma}

\begin{proofT}
	The claim follows by considering that a ${\sf read}()$ returns a value to the client after $2\delta$ time, independently of the behavior of the servers (see lines \ref{R-ITBCCAM-03}-\ref{R-ITBCCAM-06}, Figure \ref{fig:readProtocolITBCAM}).
	\renewcommand{\toto}{lem:rTermITBCAM}
\end{proofT}

\begin{theorem}[Termination]\label{th:terminationITBCAM}
	If a correct client $c_i$ invokes an operation, $c_i$ returns from that operation in finite time. 
\end{theorem}

\begin{proofT}
	The proof follows from Lemma \ref{lem:wTermITBCAM} and Lemma \ref{lem:rTermITBCAM}.
	\renewcommand{\toto}{th:terminationITBCAM}
\end{proofT}


Validity property is proved with the following steps:
\begin{itemize}
\item{1.} \op{maintenance} operation works (i.e., at the end of the operation $n-f$ servers store valid values). 
In particular, for a given value $v$ stored by $\#echo$ correct servers at the beginning of the \op{maintenance} operation, there are $n-f$ servers that may store $v$ at the end of the operation;
\item{2.} given a \op{write} operation that writes $v$ at time $t$ and terminates at time $t+\delta$, there is a time $t'>t+\delta$ after which $\#reply$ correct servers store $v$. 
\item{3.} at the next \op{maintenance} operation after $t'$ there are $\#reply-f=\#echo$ correct servers that store $v$, for step (1) this value is maintained.
\item{4.} the validity follows considering that the \op{read} operation is long enough to include the $t'$ of the last written value before the \op{read} and $V$ is big enough to do not be full filled with new values before $t'$.
\end{itemize}

%
%
%

Before to prove the correctness of the \op{maintenance} operation let us see how many Byzantine agent there may be during such operation. Since the cured server run it as soon as the mobile agent $ma_i$ leaves it, then $ma_i$ movement are aligned to such operation, this agent contribution is $\frac{2\delta}{\Delta}=k$. All the others $f-1$ mobile agent are not aligned, thus their contribution is $Max\tilde{B}(t,t+2\delta)=k+1$. Thus there are $k+(k+1)\times(f-1)$ Byzantine servers during the $2\delta$ time \op{maintenance} operation.

\begin{lemma}[Step 1]\label{lem:maintenanceNoWriteITBCAM}
	Let $T_i=t$ be the time at which mobile agent $ma_i$ leave $s_c$. Let $v$ be the value stored at $\#echo_{CAM}$  servers $s_j \notin B(t,t+\delta) \wedge s_j \in Co(t+\delta)$, $v \in V_j \forall s_j \in Co(t+\delta)$.  At time $t+2\delta$, at the end of the \op{maintenance}, $v$ is returned to $s_c$ by the function  ${\sf select\_d\_pairs\_max\_sn}(echo\_vals_c)$. 
\end{lemma}

\begin{proofL} 
	The proof follows considering that:
	\begin{itemize}
	\item the \op{maintenance} employs a request-reply pattern and during such operation, by hypothesis, there are $\#echo_{CAM}$ servers that are never affected during the $[T_i,T_i+\delta]$ time period and are correct at time $T_i+\delta$. i.e., there are $\#echo_{CAM}$ servers that deliver the \msg{echo\_req} message (the can be either correct or cured) but are correct at time $T_i+\delta$ such that the reply is delivered by $s_c$ by time $T_i+2\delta$.
	\item during the \op{maintenance} operation there are $k+(k+1)\times(f-1)$ Byzantine servers, and $(\frac{k}{2})f$ servers that were Byzantine in $[t-\delta, t]$ time period, thus they could have sent incorrect messages as well.
	\item each cured servers, invokes {\sc awareAll}$()$ function, sends a $\bot$ message twice: when they are aware to be cured and $\delta$ time after. Thus by time $t+2\delta$ server running the maintenance removes from $echo\_vals$ the $(\frac{k}{2})f$ messages sent by those servers. In the end there are $k+(k+1)\times(f-1)=(k+1)f-1$ messages coming from Byzantine servers in the $echo\_vals_c$ set. 
	\end{itemize}
	$\#echo_{CAM}=(k+1)f>(k+1)f-1$ thus Byzantine servers can not force the  ${\sf select\_d\_pairs\_max\_sn}(echo\_vals_c)$ function to return a not valid value and ${\sf select\_d\_pairs\_max\_sn}(echo\_vals_c)$ returns $v$ that occurs $\#reply_{CAM}$ times, concluding the proof.	
\renewcommand{\toto}{lem:maintenanceNoWriteITBCAM}
\end{proofL}

{


\begin{lemma}[Step 2.]\label{lem:writeCompletionITBCAM}
	Let $op_W$ be a ${\sf write}(v)$ operation invoked by a client $c_k$ at time $t_B(op_W)=t$ then at time $t+\delta$ there are at least $\#reply_{CAM}$ servers $s_j \notin B(t+\delta)$ such that $v \in V_j$. 
\end{lemma}

\begin{proofL}
The proof follows considering that during the \op{write} operation, $[t,t+\delta]$, there can be at most $(\frac{k}{2}+1)f$ mobile agents. Thus, during such time there are $n-(\frac{k}{2}+1)f=2(k+1)f+1-(\frac{k}{2}+1)f=(k+\frac{k}{2}+1)f+1$ servers $s_j$ that being either cured or correct, execute code in Figure \ref{fig:writeProtocolITBCAM}, line \ref{W-ITBSCAM-01}, inserting $v$ in $V_j$. Finally, $(k+\frac{k}{2}+1)f+1>(k+1)f+1=\#reply_{CAM}$ concluding the proof.

\renewcommand{\toto}{lem:writeCompletionITBCAM}
\end{proofL}

For simplicity, for now on, given a \op{write} operation $op_W$ we call $t_B(op_W)+\delta=t_{wC}$ the {\bf completion time} of $op_W$, the time at which there are at least $\#reply_{CAM}$ servers storing the value written by $op_W$.

\begin{lemma}[Step 3.]\label{lem:writeStableITBCAM}
Let $op_W$ be a \op{write} operation occurring at $t_B(op_W)=t$ and let $v$ be the written value and let $t_{wC}$ be its completion time. 
Then if there are no other \op{write} operations after $op_W$, the value written by $op_W$ is stored by all correct
servers forever.
\end{lemma}

\begin{proofL}
Following the same reasoning as Lemma \ref{lem:writeCompletionITBCAM}, at time $t+\delta$, assuming that in $[t,t+\delta]$ there are $(\frac{k}{2}+1)f$, then there are at least $(k+\frac{k}{2}+1)f+1$ servers $s_j$ that being either cured or correct, execute code in Figure \ref{fig:writeProtocolITBCAM}, line \ref{W-ITBSCAM-01}, inserting $v$ in $V_j$. Now let us consider the following:
\begin{itemize}
\item Let ${B}_1=\tilde{B}(t,t+\delta)$ be the set containing the $(\frac{k}{2}+1)f$ Byzantine servers during $[t,t+\delta]$, so that there are $(2k+1)f+1-\frac{k}{2}=(k+\frac{k}{2}+1)f+1\geq \#reply_{CUM}$ non faulty servers storing $v$;
\begin{itemize}
	\item  there are $(\frac{k}{2})f$ Byzantine servers in ${B}_1$ that begin the \op{maintenance} operation 
	. At that time there are $\#reply_{CAM}$ non faulty servers storing $v$, being $\#reply_{CAM}>\#echo_{CAM}$, for Lemma \ref{lem:maintenanceNoWriteITBCAM} at the end of the \op{maintenance} operation, by time $t+3\delta$, those servers obtain $v$ a result of ${\sf select\_d\_pairs\_max\_sn}(echo\_vals)$ invocation, whose is stored in $V$ since there are no other \op{write} operation and since $v$ has the highest associated sequence number.  
\end{itemize}
\item Let ${B}_2=\tilde{B}(t+\delta,t+2\delta)$ be the set containing Byzantine servers in the next $\delta$ period. Those servers are $\frac{k}{2}f$ (it is not $\frac{k}{2}f+1$, otherwise we would count the Byzantine servers at $t+\delta$ twice). Thus, at $t+2\delta$ there are $(k+\frac{k}{2}+1)f+1-\frac{k}{2}f=(k+1)f+1=\#reply_{CAM}$ non faulty servers storing $v$;
\begin{itemize}
	\item  there are $(\frac{k}{2})f$ Byzantine servers in ${B}_2$ that begin the \op{maintenance} operation during $[t+\delta,t+2\delta]$ time interval. There are $\#reply_{CAM}$ non faulty servers storing $v$, being $\#reply_{CAM}>\#echo_{CAM}$, for Lemma \ref{lem:maintenanceNoWriteITBCAM} at the end of the \op{maintenance} operation, by time $t+4\delta$, those servers, get $v$ invoking ${\sf select\_d\_pairs\_max\_sn}(echo\_vals)$, whose is stored in $V$ since there are no other \op{write} operation and since $v$ has the highest associated sequence number. 
\end{itemize}
\item Let ${B}_3=\tilde{B}(t+2\delta,t+3\delta)$ be the set containing Byzantine servers in the next $\delta$ period. Those servers are $\frac{k}{2}f$. At $t+3\delta$ there are $(k+1)f+1-\frac{k}{2}f<\#reply_{CAM}$ non faulty servers storing $v$
and the there are $(\frac{k}{2})f$ servers in ${B}_1$ that terminated the \op{maintenance} operation storing $v$. Summing up there are $(k+1)f+1-\frac{k}{2}f+\frac{k}{2}f=\#reply_{CAM}$ servers storing $v$.
\end{itemize}
Thus, after $t+3\delta$ period there are servers becoming affected that lose $v$, but there are other $f$ servers that become correct storing $v$, so that all correct servers store $v$. Since there are no more \op{write} operation, this reasoning can be extended forever, concluding the proof.
\renewcommand{\toto}{lem:writeStableITBCAM}
\end{proofL}

\begin{lemma}[Step 3.]\label{lem:permanenzaWriteITB}
Let $op_{W_{0}}, op_{W_1}, \dots, op_{W_{k-1}}, op_{W_k}, op_{W_{k+1}}, \dots$ be the sequence of \op{write} operations issued on the regular register. Let us consider a particular $op_{W_k}$, let $v$ be the value written by $op_{W_k}$ and let $t_E{w_k}$ be its completion time. Then the register stores $v$ (there are at least $\#reply_{CAM}$ correct servers storing it) up to time at least $t_B{W_{k+3}}$. 
\end{lemma}

\begin{proofL}
The proof simply follows considering that:
\begin{itemize}
\item for Lemma \ref{lem:writeStableITBCAM} if there are no more \op{write} operation then $v$, after $t_{wC}$, is in the register forever.
\item any new written value is store in an ordered set $V$ (cf. Figure \ref{fig:writeProtocolITBCAM} line \ref{W-ITBSCAM-01}) whose dimension is 3.
\item \op{write} operations occur sequentially.
\end{itemize}
It follows that after the beginning of 3 \op{write} operations, $op_{W_{k+1}},op_{W_{k+2}},op_{W_{k+3}}$, $v$ it may be no more stored in the regular register.  
\renewcommand{\toto}{lem:permanenzaWriteITB}
\end{proofL}

\begin{theorem}[Step 4.]\label{th:validityITBCAM}
	Any ${\sf read}()$ operation returns the last value written before its invocation, or a value written by a ${\sf write}()$ operation concurrent with it.
\end{theorem}

\begin{proofT}
Let us consider a \op{read} operation $op_R$. We are interested in the time interval $[t_B(op_R),$ $ t_B(op_R)+\delta]$. Since such operation lasts $2\delta$, the reply messages sent  by correct servers within  $t_B(op_R)+\delta$ are delivered by the reading client.  For $\delta \leq \Delta<3\delta$ during $[t,t+\delta]$ time interval there are $n-\frac{k}{2}-1\geq \#reply_{CAM}$ correct servers that have the time to deliver the read request and reply. Now we have to prove that what those correct servers reply with is a valid value. 
There are two cases, $op_R$ is concurrent with some \op{write} operations or not. \\
{\bf - $op_R$ is not concurrent with any \op{write} operation}. Let $op_W$ be the last \op{write} operation such that $t_E(op_W)\leq t_B(op_R)$ and let $v$ be the last written value. For Lemma \ref{lem:writeStableITBCAM} after the write completion time $t_Cw$ there are $\#reply_{CAM}$ non faulty servers storing $v$. Since $t_B(op_R)+\delta \geq t_Cw$, then there are $\#reply_{CAM}$ non faulty servers replying with $v$ (Figure \ref{fig:readProtocolITBCAM}, lines \ref{R-ITBSCAM-02}-\ref{R-ITBSCAM-03}). So the last written value is returned.\\
{\bf - $op_R$ is concurrent with some \op{write} operation}. Let us consider the time interval $[t_B(op_R),$ $ t_B(op_R)+\delta]$. In such time there can be at most two \op{write} operations. Thus for Lemma \ref{lem:permanenzaWriteITB} the last written value before $t_B(op_R)$ is still present in $\#reply_{CAM}$ non faulty servers. Thus at least the last written value is returned. \\
To conclude, for Lemma \ref{l:MaxB}, during the \op{read} operation there are at most $(k+1)f$ Byzantine servers, being $\#reply_{CAM}>(k+1)f$ then Byzantine servers may not force the reader to read another or older value and  even if an older values has $\#reply_{CAM}$ occurrences the one with the highest sequence number is chosen.  
\renewcommand{\toto}{th:validityITBCAM}
\end{proofT}

}

\begin{theorem}\label{t:fastCAMITB}
	Let $n$ be the number of servers emulating the register and let $f$ be the number of Byzantine agents in the $(ITB, CAM)$ round-free Mobile Byzantine Failure model.
	Let $\delta$ be the upper bound on the communication latencies in the synchronous system.
	If $n=n_{CAM}$ according to Table \ref{tab:summaryITBCAM} then $\mathcal{P}_{reg}$ implements a SWMR Regular Register in the $(ITB, CAM)$ and $(ITU,CAM)$ round-free Mobile Byzantine Failure model.
\end{theorem}

\begin{proofT}
	The proof simply follows from Theorem \ref{th:terminationITBCAM} and Theorem \ref{th:validityITBCAM} and considering $\Delta=1$ in the case of $(ITU,CAM)$ model.
	\renewcommand{\toto}{t:fastCAMITB}
\end{proofT}

\begin{lemma}\label{l:ITBCAMtight}
Protocol $\mathcal{P}_{reg}$ for $\delta  \leq \Delta<3\delta$ is tight with respect to $\gamma \leq 2\delta$.
\end{lemma}

\begin{proofL}
The proof follows from Theorem \ref{t:fastCAMITB} and Theorem \ref{t:LB}, i.e., upper bound and lower bound match. In particular Lower bounds are computed using the values in Table \ref{t:tabelloneGenerale} to compute $n_{CAM_{LB}}$ as defined in Table \ref{tab:summary} for $\gamma\leq 2\delta$ (cf. Lemma \ref{lem:maintenanceNoWriteITBCAM}).
\renewcommand{\toto}{l:ITBCAMtight}
\end{proofL}

\subsection{Correctness $(ITB, CUM)$}

To prove the correctness of $\mathcal{P}_{reg}$ we demonstrate that the termination property is satisfied i.e, that ${\sf read}()$ and ${\sf write}()$ operations terminates. For the validity property we follow te same four steps as defined in Section 5.1.

\begin{lemma}\label{lem:wTermITBCUM}
	If a correct client $c_i$ invokes ${\sf write}(v)$ operation at time $t$ then this operation terminates at time $t+\delta$.
\end{lemma}

\begin{proofL}
	The claim simply follows by considering that a ${\sf write\_confirmation}$ event is returned to the writer client $c_i$ after $\delta$ time, independently of the behavior of the servers (see lines \ref{W-CITBCUM-03}-\ref{W-CITBCUM-04}, Figure \ref{fig:writeProtocolITBCUM}).
	\renewcommand{\toto}{lem:wTermITBCUM}
\end{proofL}

\begin{lemma}\label{lem:rTermITBCUM}
	If a correct client $c_i$ invokes ${\sf read}()$ operation at time $t$ then this operation terminates at time $t+2\delta$.
\end{lemma}

\begin{proofL}
	The claim simply follows by considering that a ${\sf read}()$ returns a value to the client after $2\delta$ time, independently of the behaviour of the servers (see lines \ref{R-SCUM-03}-\ref{R-SCUM-06}, Figure \ref{fig:readProtocolITBCUM}).
	\renewcommand{\toto}{lem:rTermITBCUM}
\end{proofL}

\begin{theorem}[Termination]\label{th:terminationITBCUM}
	If a correct client $c_i$ invokes an operation, $c_i$ returns from that operation in finite time. 
\end{theorem}

\begin{proofT}
	The proof simply follows from Lemma \ref{lem:wTermITBCUM} and Lemma \ref{lem:rTermITBCUM}.
	\renewcommand{\toto}{th:terminationITBCUM}
\end{proofT}



To easy the next Lemmas let us use state the following result.

\begin{lemma}\label{lem:reply}
Let $[t,t+2\delta]$ be a generic interval, then there are always at least $\#reply_{CUM}$ correct servers that reply during the $[t,t+\delta]$ time interval. 
\end{lemma}

\begin{proofL}
This follows considering the definition of minimum number of correct replies during a time interval (cf. Corollary \ref{c:minimum}). Since does exist a tight protocol $\mathcal{P}$ solving a regular register in the $(\Delta S, CAM)$ model, then for Lemma \ref{l:mappingAsynchr}, is it possible to apply values from Table \ref{t:tabelloneGenerale} to compute the minimum number of correct replies during the considered time interval, substituting values in each case the result is always at least $\#reply_{CUM}$. 
\renewcommand{\toto}{lem:reply}
\end{proofL}

\begin{lemma}[Step 1.]\label{lem:maintenanceNoWriteITBCUM}
	Let $T_i$ be the time at which mobile agent $ma_i$ leave $s_c$ and let $t\leq T_i+2\delta$ the time at which $s_c$ run the second \op{maintenance} operation. Let $v$ be the value stored at $\#echo_{CUM}$  servers $s_j \notin B(t,t+\delta)$, $v \in V_j \forall s_j \notin B(t,t+\delta)$.  At time $t+2\delta$, at the end of the \op{maintenance}, $v$ is returned to $s_c$ by the function  ${\sf select\_three\_pairs\_max\_sn}(echo\_vals_c)$.  
\end{lemma}

\begin{proofL} 
	The proof follows considering that:
	\begin{itemize}
	\item the \op{maintenance} employs a request-reply pattern and during such operation, by hypothesis, there are $\#echo_{CUM}$ servers that are never affected during the $[t,t+\delta]$ time period and are storing $v$ at time $t+\delta$. i.e., there are $\#echo_{CUM}$ servers that deliver the \msg{echo\_req} message (the can be either correct or cured) but are storing $v$ in $V$ at time $t+\delta$ such that the reply is delivered by $s_c$ by time $t+2\delta$.
	\item during the \op{maintenance} operation can incorrectly contribute $(k+1)f$ Byzantine servers, and $(2k)f$ servers that were Byzantine in $[t-4\delta, t]$ time period, thus they could be still in a cured state \footnote{We prove hereafter that $\gamma \leq 4\delta$, but to prove it we have first to prove that the \op{maintenance} lasts $2\delta$ time.}.
	\item when the \msg{echo\_req} message is sent, $s_c$ uses a random number in order to be able to accept only \msg{echo} message sent after $t$. 
	\end{itemize}
	$\#echo_{CUM}=(3k)f+1>3kf$ thus Byzantine servers can not force the  ${\sf select\_three\_pairs\_max\_sn}(echo\_vals_c)$ function to return a not valid value so it returns $v$ that occurs $\#reply_{CUM}$ times, which is true since there exist $\#echo\_{CUM}$ non faulty servers that reply to the \msg{echo\_req} message sending back $v$, concluding the proof.	
\renewcommand{\toto}{lem:maintenanceNoWriteITBCUM}
\end{proofL}

In the sequel we consider $\gamma\leq 4\delta$. In the previous Lemma we proved that cured servers $s_c$ can get valid values in $2\delta$ time. Contrarily to all the previous model, the \op{maintenance} operation is triggered each $2\delta$. Thus a mobile agent, just before to leave could leave $s_c$ with the timer just reset and garbage in the $echo\_set_c$ and $V_c$ sets, which does not allow $s_c$ to correctly terminate the operation. Thus $s_c$ has to wait $2\delta$ before to effectively starts a correct \op{maintenance} operation. In the sequel we refer to the {\bf first maintenance} as the operation that may be ineffective and we refer to the {\bf second maintenance} as the operation that allows a cured server to retrieve and store valid values. It is straightforward that $\gamma\leq4\delta$ and the next Corollary just follows.

\begin{corollary}\label{c:gammaITBCUM}
Protocol $\mathcal{P}$ implements a \op{maintenance} operation that implies $\gamma \leq 4\delta$.
\end{corollary}


\begin{lemma}[Step 2.]\label{lem:writeCompletionITBCUM}
	Let $op_W$ be a ${\sf write}(v)$ operation invoked by a client $c_k$ at time $t_B(op_W)=t$ then at time $t+\delta$ there are at least $n-2f>\#reply_{CUM}$ non faulty servers $s_i$ such that $v \in W_i$ (so that when $s_i$ invokes ${\sf conCut}(V_i, V_{safe_i}, W_i)$ $v$ is returned).
\end{lemma}

\begin{proofL}
When the \msg{write} message is delivered by non faulty servers $s_i$, such message is stored in $W_i$ and a timer associated to it is set to $4\delta$, after that the value expires. For Lemma \ref{l:MaxB} in the $[t,t+\delta]$ time interval there are maximum $2f$ Byzantine servers. All the remaining $n-2f$ non faulty servers execute the correct protocol code, Figure \ref{fig:writeProtocolITBCUM} line \ref{W-SITBCUM-01} inserting $v$ in $W_i$. Since \op{write} operations are sequential, during $[t,t+\delta]$ there is only one new value inserted in $W_i$, which is returned by the function ${\sf conCut}()$ by construction.  
\renewcommand{\toto}{lem:writeCompletionITBCUM}
\end{proofL}

For simplicity, for now on, given a \op{write} operation $op_W$ we call $t_B(op_W)+\delta=t_{wC}$ the {\bf completion time} of $op_W$, the time at which there are at least $\#reply_{CUM}$ servers storing the value written by $op_W$.

\begin{lemma}[Step 3.]\label{lem:writeStableITBCUM}
Let $op_W$ be a \op{write} operation and let $v$ be the written value and let $t_{wC}$ be its time completion. 
Then if there are no other \op{write} operation, the value written by $op_W$ is stored by all correct servers forever (i.e., $v \in {\sf conCut}(V_i, V_{safe_i}, W_i)$).
\end{lemma}

\begin{figure*}
	\begin{tikzpicture}[y=-1cm]
		\def \lenght {12.1}
		\def \n {13} 
		
		\def \deltaGrande {.5}
		\def \deltaPiccolo {.5}
		\def \gammaCuring {2}
		
		\processes
		\faults{0}{0}{20}{yellow}{3}
		\curatoParziale{21*\deltaGrande}{6}{\lenght-21*\deltaGrande-0.1}{yellow}
		\node[] () at (22*\deltaGrande,6.7) {$\dots$};
		
		\lineaVerticale{0.2}{-1}{\footnotesize {\sf write}$(v)$}{}
		\lineaVerticale{2.2}{-1}{$t'$}{}
		\punto{2.2}{5}{green}{$v\in V_{safe_5}$}
		\punto{2.2}{6}{green}{$v\in V_{safe_6}$}
		\punto{2.2}{7}{green}{$v\in V_{safe_7}$}
		\punto{2.2}{8}{green}{$v\in V_{safe_8}$}
		\punto{2.2}{9}{green}{$v\in V_{safe_9}$}
		\punto{2.2}{9.5}{green}{$v\in V_{safe_{10}}$}
		\punto{2.2}{10.5}{green}{$v\in V_{safe_{11}}$}
		\punto{2.2}{12}{green}{$v\in V_{safe_{12}}$}
		\punto{2.2}{13}{green}{$v\in V_{safe_{13}}$}
		\lineaVerticale{2.7}{-1}{}{dotted}
		\lineaVerticale{3.2}{-1}{$t'+2\delta$}{}
		\punto{2.5}{0}{green}{$v\in V_{safe_0}$}
		\punto{3}{1}{green}{$v\in V_{safe_1}$}
		
		\rettangoloOpaco{1.6}{10}{1}{white}\rettangoloOpaco{2.6}{10}{1}{white}
		\rettangoloOpaco{2}{11}{1}{white} \rettangoloOpaco{3}{11}{1}{white}
		\node[] () at (2.1,9.9){\scriptsize $op_{M_{-1}}$}; \node[] () at (3.1,9.9){\scriptsize $op_{M_1}$};
		\node[] () at (2.5,10.9){\scriptsize $op_{M_{-1}}$}; \node[] () at (3.5,10.9){\scriptsize $op_{M_1}$};

	\end{tikzpicture}
	\caption{\op{maintenance} operation $op_{M_1}$ analysis after a \op{write} operation, $t'=t+4\delta$. White rectangles are \op{maintenance} operation run by correct servers. In particular $s_{10}$ runs such operation during the first $\delta$ period after $t'$, while $s_11$ runs it during the second $\delta$ period.}\label{fig:maintenanaceITBCUM}
\end{figure*}
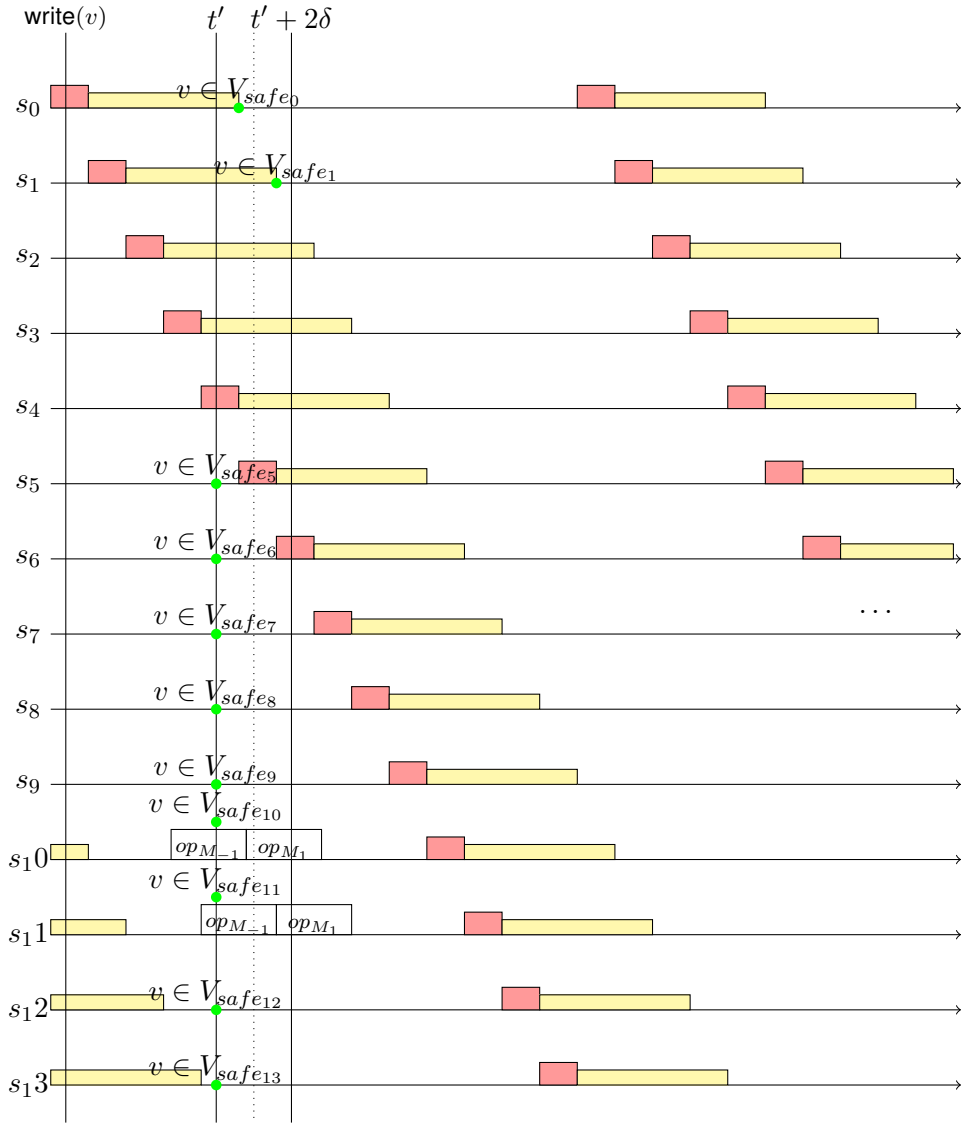

\begin{proofL}
From Lemma \ref{lem:writeCompletionITBCUM} at time $t_{wC}$ there are at least $n-2f>\#reply_{CUM}$ non faulty servers $s_j$ such that $v \in W_i$. 
For sake of simplicity let us consider Figure \ref{fig:maintenanaceITBCUM}. Let us consider that:
\begin{itemize}
\item for Lemma \ref{lem:writeCompletionITBCUM}, all non faulty servers $s_i$ have $v$ in $W_i$ at most at $t_{wC}$;
\item when $s_i$ runs the next \op{maintenance}, $v$ is returned by ${\sf select\_three\_pairs\_max\_sn}(echo\_vals_i)$ function at the end of such operation, and since it is the value with the highest sequence number (there are no other \op{write} operation) then $v$ is inserted in $V_{safe_i}$  (cf. Figure \ref{fig:stateMaintenanceProtocolITBCUM} line \ref{M-ITBCUM-06}), thus such value is present in the \msg{ECHO} message replies for the next $2\delta$ time;
\item this is trivially true up to time $t'=t+4\delta$, for the timer associated to each $v$ in $W_i$. In $[t,t']$ there are $2k+1$ Byzantine servers, thus $v \in W_j$ at $n-(2k+1)$ non faulty servers, and  $n-(2k+1)=(3k+1)f+1=\#reply_{CUM}\geq\#echo_{CUM}$;
\item for each non faulty server the next \op{maintenance} operation $op_M$ can happen either in $[t', t'+\delta]$ or in $[t'+\delta, t'+2\delta]$ (cf. Figure \ref{fig:maintenanaceITBCUM})$s_{10}$ and $s_{11}$ respectively:
\begin{itemize}
\item{$t_B(op_M)\in[t', t'+\delta]$ (cf. $s_{10}$ Figure \ref{fig:maintenanaceITBCUM})}: $s_{10}$ starts $op_{M_1}$ before $t'+\delta$, let us name it server type A. This means that $t_B(op_{M_{-1}})+\delta<t'-\delta$, thus for Lemma \ref{lem:maintenanceNoWriteITBCUM}, at the end of the operation $v \in V_{safe_{10}}$ and during $op_{M_1}$ $v \in V_{10}$;
\item{$t_B(op_M)\in[t'+\delta, t'+2\delta]$ (cf. $s_{11}$ Figure \ref{fig:maintenanaceITBCUM})}: $s_{11}$ starts $op_{M_1}$ after $t'+2\delta$ let us name it server type B. This means that $t_B(op_{M_{-1}})+\delta>t'$, thus at the end of the operation we can not say that $v \in V_{safe_10}$ but at least during $op_{M_{-1}}$ $v \in V_{11}$.
\end{itemize}
\end{itemize}
If all non faulty servers are type A, during $op_{M_1}$ all non faulty servers have $v \in V$ and insert $v$ in the \msg{echo} message. 
The same happens if all non faulty servers are type B, during $op_{M_{-1}}$, all of them inter $v$ in the \msg{echo} message and the \op{maintenance} operation terminates with such value. If the situation is mixed, then servers type B, when run $op_{M_{-1}}$, deliver \msg{echo} messages from both type A and type B servers. Thus if there are enough occurrence of $v$ they can store $v \in V_{safe_b}$ and during $op_{M_{1}}$ $v \in V_b$. During such operation both servers type A and type B have $v in V$. Again,  if there are enough occurrences of $v$, the operation ends with $v \in V_{safe_b}$. It follows that servers type A, when run $op_{M_1}$ delivers \msg{echo} messages containing $v$ from both type A and type B servers. During the time interval $[t',t'+2\delta]$ there are $k$ correct servers that are affected by mobile agent, cf. Figure \ref{fig:maintenanaceITBCUM}, $s_5$ and $s_6$. At the same time there is server $s_0$, type A, that terminate its \op{maintenanace} with $v\in V_{safe_0}$, and thus compensates $s_5$, allowing $s_1$, type B, to terminate the \op{maintenanace} operation with $v \in V_{safe_1}$, which compensates $s_6$. This cycle, between type A and type B servers can be extended forever. By hypothesis there are no more \op{write} operation, thus all correct servers have $v\in V_{safe}$ or $V$, and $v$ is returned when servers invoke function  ${\sf conCut}()$.  
\renewcommand{\toto}{lem:writeStableITBCUM}
\end{proofL}

\begin{lemma}[Step 3.]\label{lem:permanenzaWriteITBCUM}
Let $op_{W_{0}}, op_{W_1}, \dots, op_{W_{k-1}}, op_{W_k}, op_{W_{k+1}}, \dots$ be the sequence of \op{write} operation issued on the regular register. Let us consider a generic $op_{W_k}$, let $v$ be the written value by such operation and let $t_{wC}$ be its completion time. Then $v$ is in the register (there are $\#reply_{CUM}$ correct servers that return it when invoke the function ${\sf conCut}()$) up to time at least $t_B{W_{k+3}}$. 
\end{lemma}

\begin{proofL}
The proof simply follows considering that:
\begin{itemize}
\item for Lemma \ref{lem:writeStableITBCUM} if there are no more \op{write} operation then $v$, after $t_{wC}$, is in the register forever.
\item any new written value eventually is stored in an ordered set $V_{safe}$ and then $V$(cf. Figure \ref{fig:stateMaintenanceProtocolITBCUM} line \ref{M-ITBCUM-01} or line \ref{M-ITBCUM-06}) whose dimension is three.
\item \op{write} operation occur sequentially.
\end{itemize}
It follows that after three \op{write} operations, $op_{W_{k+1}},op_{W_{k+2}},op_{W_{k+3}}$ in $V$ $V_{safe}$ and $W$ there are three values whose sequence number is higher than the one associated to $v$, thus by construction ${\sf conCut}()$ does not return $v$ anymore, $v$ is no more stored in the regular register. 
\renewcommand{\toto}{lem:permanenzaWriteITBCUM}
\end{proofL}

\begin{theorem}[Step 4.]\label{th:validityITBCUM}
	Any ${\sf read}()$ operation returns the last value written before its invocation, or a value written by a ${\sf write}()$ operation concurrent with it.
\end{theorem}

\begin{proofT}
Let us consider a \op{read} operation $op_R$. We are interested in the time interval $[t_B(op_R),t_B(op_R)+\delta]$. Since such operation lasts $2\delta$, the reply messages sent  by correct servers within  $t_B(op_R)+\delta$ are delivered by the reading client.  During $[t,t+\delta]$, for Lemma \ref{lem:reply} there are at least $\#reply_{CUM}$ correct servers that reply. Now we have to prove that what those correct servers reply with is a valid value.
There are two cases, $op_R$ is concurrent with some \op{write} operations or not. \\
{\bf - $op_R$ is not concurrent with any \op{write} operation}. Let $op_W$ be the last \op{write} operation such that $t_E(op_W)\leq t_B(op_R)$ and let $v$ be the last written value. For Lemma \ref{lem:writeStableITBCUM} after the write completion time $t_{wC}$ there are at least $\#reply_{CUM}$ correct servers storing $v$ (i.e., $v \in {\sf conCut}(V_j, V_{safe_j})$). Since $t_B(op_R)+2\delta \geq t_Cw$, then there are $\#reply_{CUM}$ correct servers replying with $v$ (cf. Lemma \ref{lem:reply}), by hypothesis there are no further \op{write} operation and $v$ has the highest sequence number. It follows that the last written value $v$ is returned.\\
{\bf - $op_R$ is concurrent with some \op{write} operation}. Let us consider the time interval $[t_B(op_R), t_B(op_R)+\delta]$. In such time there can be at most two \op{write} operations. Thus for Lemma \ref{lem:permanenzaWriteITBCUM} the last written value before $t_B(op_R)$ is still present in $\#reply_{CUM}$ correct servers and all of them reply (cf. Lemma \ref{lem:reply}) thus at least the last written value is returned. 
To conclude, for Lemma \ref{l:MaxB}, during the \op{read} operation there are at most $(k+1)f$ Byzantine servers and $2k$ cured servers \footnote{Servers where affected in the previous $4\delta$ time period, thus they are still running the two \op{maintenance} operations, that last at most $4\delta$.}, being $\#reply_{CUM}=(3k+1)f+1>(3k+1)f$ then Byzantine servers may not force the reader to read another or older value and  even if an older values has $\#reply_{CUM}$ occurrences the one with the highest sequence number is returned, concluding the proof.  
\renewcommand{\toto}{th:validityITBCUM}
\end{proofT}

\begin{theorem}\label{t:fastITBCUM}
	Let $n$ be the number of servers emulating the register and let $f$ be the number of Byzantine agents in the $(ITB, CUM)$ round-free Mobile Byzantine Failure model.
	Let $\delta$ be the upper bound on the communication latencies in the synchronous system.
	If $n \ge (5k + 2)f + 1$, then $\mathcal{P}_{reg}$ implements a SWMR Regular Register in the $(ITB, CUM)$ round-free Mobile Byzantine Failure model.
\end{theorem}

\begin{proofT}
	The proof simply follows from Theorem \ref{th:terminationITBCUM} and Theorem \ref{th:validityITBCUM}.
	\renewcommand{\toto}{t:fastITBCUM}
\end{proofT}

\begin{lemma}\label{l:ITBCUMtight}
Protocol $\mathcal{P}_{reg}$ is tight in the $(ITB, CUM)$ model with respect to $\gamma \leq 4\delta$.
\end{lemma}

\begin{proofL}
The proof follows from Theorem \ref{t:fastITBCUM} and Theorem \ref{t:LB}, i.e., upper bound and lower bound match. In particular Lower bounds are computed using the values in Table \ref{t:tabelloneGenerale} to compute $n_{CUM_{LB}}$ as defined in Table \ref{tab:summary} for $\gamma\leq 4\delta$ (cf. Corollary \ref{c:gammaITBCUM}).
\renewcommand{\toto}{l:ITBCUMtight}
\end{proofL}

\section{Concluding remarks}\label{sec:conclusion}


We proposed  lower bounds and matching upper bounds for the emulation of a regular register in the round free synchronous communication model  under \emph{unsynchronized} moves of Byzantine agents. The computed lower bounds are significantly higher than those computed for  \emph{synchronized} Byzantine agents model.  Investigating other classical problems  in the same fault model is a challenging path for future research.



\bibliography{references,references1,biblio}

\end{document}